\documentclass[preprint]{elsarticle}
\usepackage[total={450pt, 595pt}]{geometry}
\usepackage{setspace}
\usepackage{etoolbox}
\usepackage{amsmath}
\usepackage{amsthm}
\usepackage{amssymb}
\usepackage{amsfonts}
\usepackage{booktabs}
\usepackage{algorithmic}
\usepackage{algorithm}
\usepackage{graphicx}
\usepackage{color}
\usepackage[tight]{subfigure}
\usepackage{threeparttable}
\usepackage{mathrsfs}
\usepackage{multirow}
\usepackage{microtype}
\usepackage{stmaryrd}
\newtheorem{theorem}{Theorem}

\usepackage{hyperref}

\begin{document}

\begin{frontmatter}

\title{A Black-box Adversarial Attack Strategy with Adjustable Sparsity and Generalizability for Deep Image Classifiers}

\author[add6]{Arka Ghosh}
   \ead{arka.ghosh@bennett.edu.in}
   \author[add2]{Sankha Subhra Mullick}
   \ead{mullicksankhasubhra@gmail.com }
    \author[add3]{Shounak Datta}
    \ead{shounak.jaduniv@gmail.com }
   \author[add2]{Swagatam Das\corref{cor1}}
   \ead{swagatam.das@isical.ac.in}
   \author[add4]{Asit Kr. Das}
   \ead{akdas@cs.iiests.ac.in}
    \author[add5]{Rammohan Mallipeddi}
   \ead{mallipeddi.ram@gmail.com }
   \cortext[cor1]{Please address correspondence to Author 4}
   \address[add1]{School of Engineering and Applied Sciences, Bennett University, India}
   \address[add2]{Electronics and Communication Sciences Unit, Indian Statistical Institute, Kolkata, India}
   \address[add3]{Electrical and Computer Engineering Dept., Duke University, USA}
   \address[add4]{Dept. of Computer Science \& Technology, Indian Institute of Engineering Science \& Technology, Shibpur, India}
   \address[add5]{College of IT Engineering, Kyungpook National University, Republic of Korea}
   \address[add6]{School of Engineering \& Applied Sciences, Bennett University,  Greater Noida, India }

\begin{abstract}
Constructing adversarial perturbations for deep neural networks is an important direction of research. Crafting image-dependent adversarial perturbations using white-box feedback has hitherto been the norm for such adversarial attacks. However, black-box attacks are much more practical for real-world applications. Universal perturbations applicable across multiple images are gaining popularity due to their innate generalizability. There have also been efforts to restrict the perturbations to a few pixels in the image. This helps to retain visual similarity with the original images making such attacks hard to detect. This paper marks an important step that combines all these directions of research. We propose the DEceit algorithm for constructing effective universal pixel-restricted perturbations using only black-box feedback from the target network. We conduct empirical investigations using the ImageNet validation set on the state-of-the-art deep neural classifiers by varying the number of pixels to be perturbed from a meager 10 pixels to as high as all pixels in the image. We find that perturbing only about 10\% of the pixels in an image using DEceit achieves a commendable and highly transferable Fooling Rate while retaining the visual quality. We further demonstrate that DEceit can be successfully applied to image-dependent attacks as well. In both sets of experiments, we outperform several state-of-the-art methods.
\end{abstract}

\begin{keyword}
Adversarial attack, black-box attack, ImageNet, deep neural networks .
\end{keyword}

\end{frontmatter}

\section{Introduction}
\label{intro}

The remarkably efficient Convolutional Neural Networks (CNN) are susceptible to perturbations in the input \cite{szegedy2013intriguing} akin to the vulnerable heel of the mighty Achilles. As CNNs are being increasingly deployed as classifiers, object detectors etc. in critical applications like autonomous driving, fraud detection, etc. their sensitivity to perturbations can result in serious adversity \cite{Li2021handwritten}. This inspired the research community to consider the study of adversarial perturbations for deep convolutional image classifiers \cite{moosavi2016deepfool} and object detectors \cite{Xiao2021Object} as immensely important and subsequently venture diverse schools of thought.

Adversarial perturbations \cite{Biggio2018Wild} can be devised using two main strategies, namely white-box and black-box attacks. In the initially designed and thus more widely explored white-box setting \cite{goodfellow2014explaining,carlini2017towards}, the attacker has full access to the parameters, the gradients of the loss function concerning the target class label(s) to the input image(s), and the architecture of the CNN which is to be attacked. However, model-specific information is likely to be often unavailable in real-world settings, thus a more practical problem was set in the form of black-box attack \cite{papernot2017practical}. In the black-box setting contrary to the white-box the access of the attacker is restricted to only the decisions made by the target classifier. Since black-box approaches \cite{dang2017evading} do not rely on any classifier-intrinsic information, the resulting perturbations generally have greater transferability compared to white-box attacks. Moreover, by design black-box setting allows employing a wider variety of strategies such as ensemble to device better performing attacks \cite{Hang2020ensemble}. 

Adversarial attacks on CNNs can also be classified as either image-dependent or universal. Image-Dependent Attacks (IDAs) \cite{Zhao_2019_ICCV} allow distinct perturbations to be employed for different images, resulting in a significant loss of the classification accuracy for the target CNN. In contrast, Universal Adversarial Attacks (UAAs) \cite{moosavi2017universal} attempt to craft perturbations that can be applied to any natural image (irrespective of content and provenance) to mislead the target classifier. As a result, UAAs are likely to exhibit better generalizability to previously unseen images \cite{Li2021object}. Moreover, UAAs are also more computationally efficient as there is no need to recompute perturbations for every individual image.

Another emerging direction of research investigates the potential for misleading the CNN classifiers by perturbing a limited number of pixels in the images \cite{shiva2017simple}. Su \textit{et al.} \cite{su2019one}  attempted to test the extent of such sparse perturbations and found that manipulating merely a single pixel can result in a successful IDA. This suggests that identifying and perturbing the pixels which are most critical for classification (as opposed to perturbing the entire image) may be sufficient to induce misclassification. Further, limiting the number of pixels to be perturbed generally results in perturbations that are imperceptible to human eyes. Such attacks are especially menacing as they can pass undetected even if the autonomous system is accompanied by human users. 

\begin{figure}[!ht]
    \centering
    \subfigure[\label{fig:motivRes1}]{\includegraphics[width=0.35\linewidth]{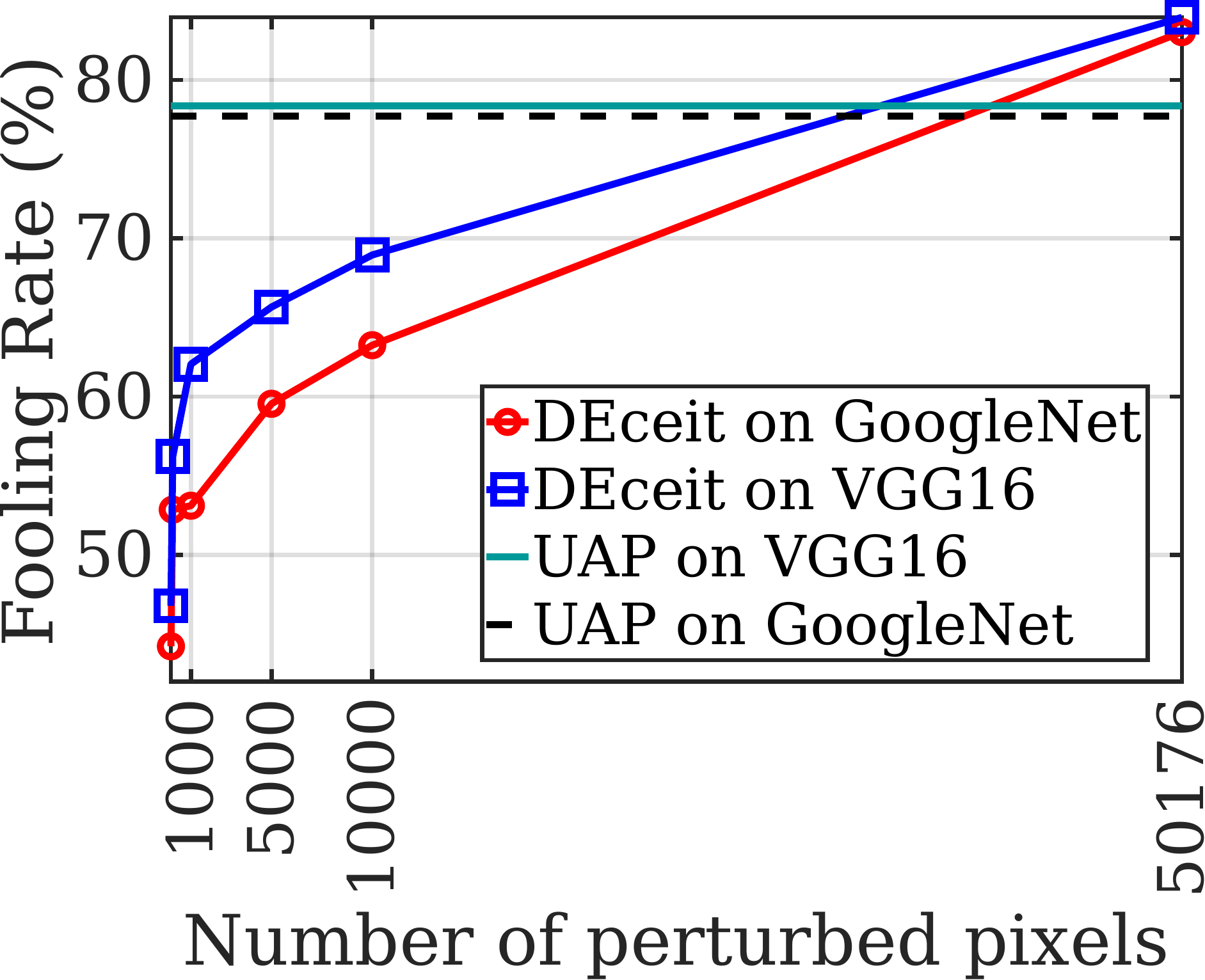}} \hspace{30pt}
    \subfigure[\label{fig:motivRes2}]{\includegraphics[width=0.35\linewidth]{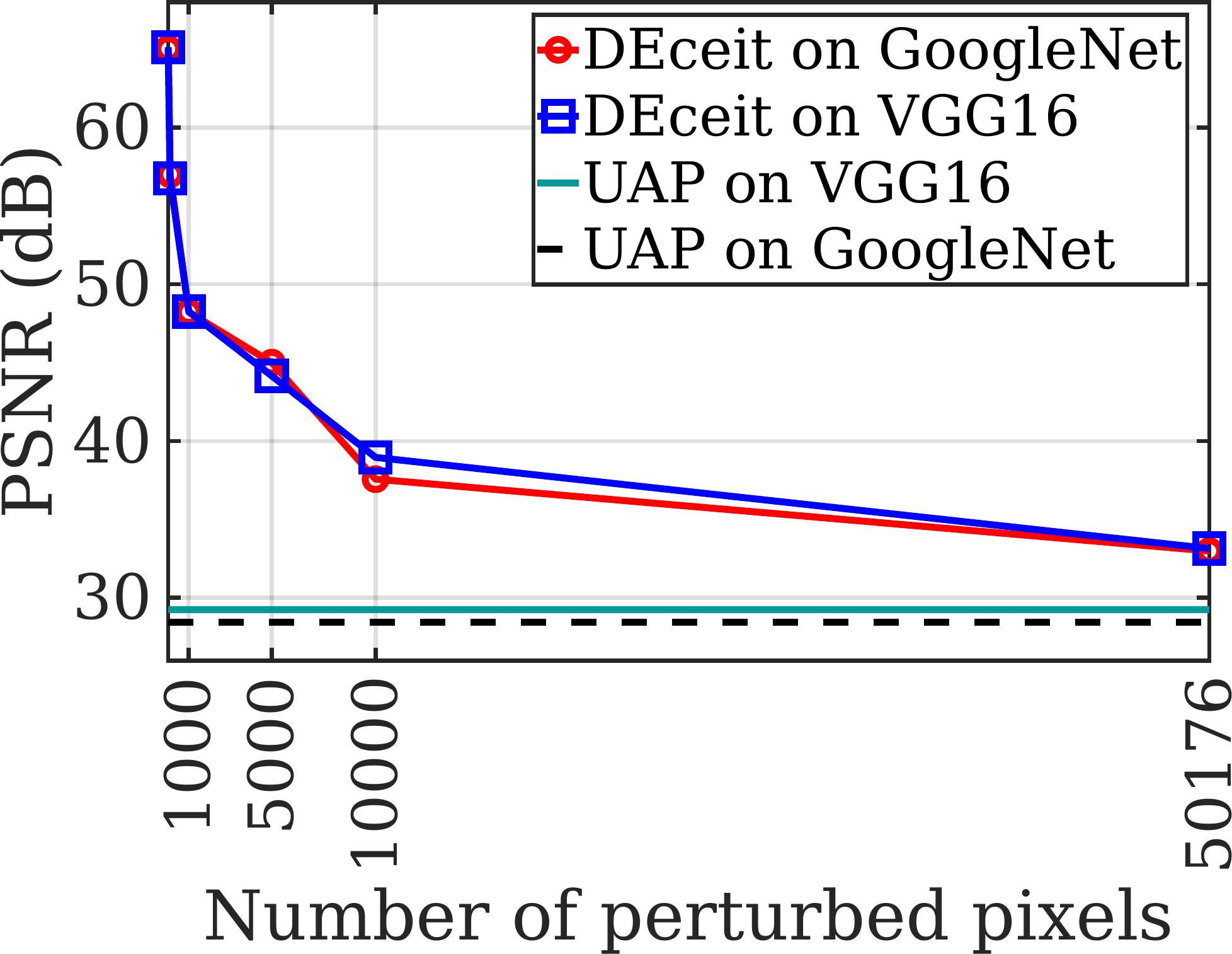}}
    \caption{We compare the quality of universal adversarial perturbations $\Delta$ generated by DEceit with increasing number of perturbed pixels $\rho$ (respectively 10, 100, 1000, 5000, 10000, and all of 50176 pixels in images of resolution $224 \times 224$) on the ImageNet 2012 validation set, for VGG16 and GoogleNet. To compensate for the extremely low number of pixels, the allowable distortion is set to [-128, 128] for 10 pixels, [-30, 30] for 100 pixels, [-20, 20] for 1000 pixels and the conventional [-10, 10] thereafter ([$a,b$] denotes closed interval of real numbers between $a$ and $b$). (a) When all pixels are allowed to be perturbed, DEceit achieves a better Fooling Rate (the percentage of test samples that alter the prediction of the target classifier after applying perturbation) than the white-box UAP \protect\cite{moosavi2017universal} on both classifiers. Moreover, DEceit achieves commendable Fooling Rate of 66\% and 60\% respectively on VGG16 and GoogleNet when only 5000 pixels are allowed to be perturbed. This indicates that it is indeed feasible to perform an effective UAA by only modifying a restricted number of pixels. (b) DEceit also achieves better PSNR values indicating a higher similarity of the perturbed image with the corresponding original.}
    \label{fig:motivRes}
\end{figure}

In light of the above discussion, in this article, we attempt for the first time, to devise pixel-restricted i.e. sparse perturbations for universal attacks by using black-box feedback. The pixel restriction means that we must identify the most critical pixels among a potentially large number of pixels, depending on the size of the input images. Moreover, the optimization problem by nature is extremely high-dimensional (for example, as described in Section \ref{proposed} given an RGB image of resolution $224 \times 224$, in the case of the proposed method the dimension of the problem may reach a maximum of 250880 when all pixels are allowed to be perturbed). Consequently, a brute-force search of the whole search space is not computationally feasible. Moreover, the black-box setting means that no gradient information is available from the target CNN, ruling out the use of gradient-based optimization schemes. To address this issue, we propose to use a powerful population-based stochastic optimization technique called Differential Evolution (DE) \cite{storn1997differential} to find suitable perturbations. DE is remarkably efficient over diverse optimization scenarios like constrained, multi-modal, multi-objective problems, owing to the decades of research effort devoted to improving its scalability, robustness \cite{das2016recent}, and convergence \cite{opara2019}. Recently, SwitchDE \cite{ghosh2017switched} showed that simple parameter and mutation switching can be very effective for high-dimensional optimization problems by achieving an effective trade-off between the exploratory and exploitative search behaviors of DE. In Section \ref{proposed} our proposed DE-based adversarial perturbation generation scheme, DEceit combines the mutation switching strategy from SwitchDE with uniform random scale-factor switching to effectively solve the high-dimensional optimization problem. The effectiveness of DEceit in comparison to UAP \cite{moosavi2017universal} is illustrated in Figure \ref{fig:motivRes}, which validates that even by perturbing merely 10\% of the total number of pixels DEceit can achieve an admirable Fooling Rate (the fraction of test samples that are differently labeled after being perturbed) on both CNNs while maintaining a better visual quality. The principal highlights of our proposal are in order, which we also illustrate in Figure \ref{fig:contrib}:

\begin{figure}[!t]
    \centering
    \includegraphics[width=0.85\textwidth]{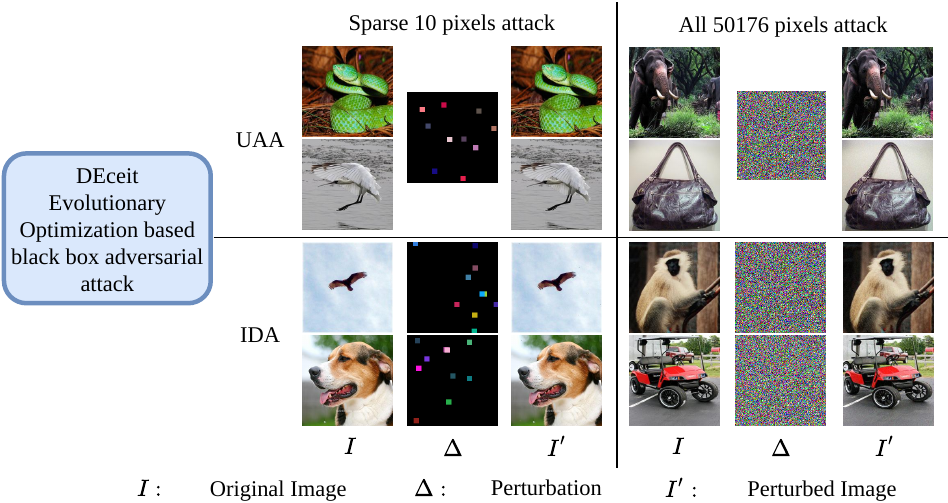}
    \caption{We present the output of the proposed DEceit on 8 images (from the ImageNet 2012 validation set) all of which successfully fooled the VGG16 after perturbation. Our technique is capable of four different types of attacks. Firstly, DEceit can perform a sparse UAA by only modifying as less as 10 pixels. Secondly, DEceit can perform UAA by perturbing all pixels (in case of an image having $224 \times 224$ resolution this can be as large as 50176). DEceit can also perform IDA by generating sparse perturbations which perturb only 10 pixels. Finally, DEceit can also generate perturbations that modify all the 50176 pixels to successfully perform an IDA. 
    To the best of our knowledge, this is the first technique that attempts to efficiently handle such a large set of diverse adversarial attack scenarios. In this example, for sparse perturbations, the range of modification is kept between [-128, 128] for the 10-pixel perturbations while the same is reduced to [-10, 10] when all pixels are allowed to be perturbed. The perturbed images generated by DEceit offer minimal visual distortions while being capable of fooling the deep CNN classifier. In this figure, the perturbed pixels for the sparse attacks have been magnified to increase visibility.} 
    \label{fig:contrib}
\end{figure}

\begin{enumerate}
    \item In Section \ref{proposed} we add to the existing literature (discussed in Section \ref{background}) by proposing DEceit, which is capable of crafting effective universal perturbations by only modifying a small fraction of the pixels for any image using black-box feedback from the network under attack. This combines the practicality of the black-box setting with the robustness of UAAs while limiting the distortions in perturbed images, making detection and prevention challenging. 
    \item Unlike Su \textit{et al.} and Alzanot \textit{et al.} \cite{alzanot2019genattack} we consider the large-scale and high-dimensional nature of the optimization problem related to adversarial attacks. Hence, in DEceit we augment DE with a couple of fundamental changes, namely we introduce a random parameter switching strategy in addition to a success-based mutation strategy adopted from Ghosh \textit{et al}. \cite{ghosh2017switched}. 
    \item We demonstrate that DEceit can directly be used to construct effective image-dependent perturbations as well.
    \item In Section \ref{sec:exp} we carry out extensive comparisons with state-of-the-art techniques to establish the ability of DEceit to generate both sparse and non-sparse perturbations which can effectively perform black-box UAA and IDA. 
\end{enumerate}

\section{Related Works}\label{background}
The work of Narodytska and Kasiviswanathan \cite{shiva2017simple} has recently been carried forward by research works like \cite{su2019one,Modas_2019_CVPR,Croce_2019_ICCV} that indicated that sparse perturbations are not only effective but are also likely to offer better imperceptibility. However, these notable attempts for generating sparse and efficient adversarial perturbation have only been applied for IDAs. In the case of UAAs, the complexity of the problem considerably increases as it becomes even more difficult to find a limited set of pixels, modifying which will affect a large set of images containing a variety of objects in differing scales, lighting conditions, etc. 

Evolutionary Optimization (EvO) algorithms gained popularity in devising adversarial perturbations as they can handle large-scale problems and are especially useful in the black-box setting. A couple of works \cite{su2019one,alzanot2019genattack} attempted to respectively utilize DE and Genetic Algorithm (GA) to carry out successful IDA. However, as discussed earlier in Section \ref{intro}, the underlying optimization problem is extremely high-dimensional in nature. A fact which Alzanot \textit{et al.}  \cite{alzanot2019genattack} did not consider and Su \textit{et al.} \cite{su2019one} avoided by restricting themselves to one pixel attack. This led them to rely on the canonical variants of the EvO algorithms which are susceptible to the curse of dimensionality alongside being ineffective in a large-scale optimization problem. 

\section{Proposed Method}\label{proposed}
An image $I$ is conventionally represented in the form of a third-order tensor i.e. $I \in \mathbb{R}^{w \times h \times c}$, where $w$, $h$ and $c$ are positive integers respectively denoting the width, height and the number of channels of the image. Depending on the application, the components of $I$ can be further constrained and bounded. Here, for simplicity without loss of generality, we assume any component $I_{pqr} \in \llbracket 0, 255 \rrbracket$, for all $p \in \llbracket 1, w \rrbracket$, $q \in \llbracket 1, h \rrbracket$, and $r \in \llbracket 1, c \rrbracket$, where $\llbracket a, b \rrbracket$ denotes the closed interval of integers between $a$ and $b$. Further, let a dataset $\mathcal{S}=\{I_{1}, I_{2}, \cdots, I_{m}\}$ be a collection of $m$ images, where each image can be classified into one of the $C$ predefined classes. Classification can be represented as finding the many-to-one mapping $\hat{y}: \mathcal{S} \rightarrow \mathcal{C}$, where $\mathcal{C}=\{1, 2, \cdots, C\}$. CNN classifiers, being supervised learning algorithms, attempt to approximate $\hat{y}$ as $y$, utilizing the information provided through a set $\mathcal{X} \subseteq \mathcal{S}$ of training examples such that for all $I \in \mathcal{X}$ the true class label $\hat{y}(I)$ is known in advance. 

\subsection{Universal Attack}
The vulnerability of a trained CNNs to adversarial attacks may originate from its learning mechanism. The decision strategy learned by the network to recognize a class can be effectively exploited by a perturbation leading to a misclassification \cite{jetley2018}. This indicates the possibility of a UAA exploiting an amalgamation of the network's decision strategies for the individual classes. 

Given an image $I \in \mathcal{S}$ a perturbed image is created as $\bar{I}=\psi(I+\Delta)$, where $\Delta \in \mathbb{R}^{w \times h \times c}$ is a perturbation tensor and $\psi(\cdot)$ is a function ensuring (commonly by scaling and/or clipping) that the addition adheres to the properties of an image. However, even a properly trained CNN may not always correctly classify an unperturbed training image, i.e. predict the label of an input $I \in \mathcal{X}$ as $y(I)$ where $y(I) \neq \hat{y}(I)$. Further, the original label $\hat{y}(I)$ of an input $I$ may not always be known in advance. Thus, considering the original labels for assessing the quality of an adversarial attack may not be useful in practice. Hence, it suffices to check if the prediction for an unperturbed image $I$ varies from the labelling of its perturbed version $\bar{I}$ by a CNN i.e. only compare $y(I)$ and $y(\bar{I})$. Our objective of designing an UAA can thus be reduced to finding a perturbation $\Delta$, which maximizes the probability of alternatively classifying $\bar{I}$:
\begin{equation}
\label{eqn:opt1}
    \max \mathcal{P} (\Delta) = \sum_{I \in \mathcal{X}} Pr({y}(\psi(I+\Delta)) \neq y(I)),
\end{equation}
\begin{equation}
    \text{s.t.} \; -\beta \leq \Delta_{pqr} \leq \beta,
\end{equation}
where $\beta$ is an user-defined parameter controlling the extent of distortion in a single pixel, and $Pr(\cdot)$ denotes the probability of an event.

\subsection{Sparse Perturbation}
Perturbing every pixel of an image may lead to compromised visual quality. An attractive alternative may come in the form of sufficient perturbation of a limited number of critical pixels. Such a perturbation will likely be capable of performing a successful attack alongside improved imperceptibility. This would enable a sparse UAA, consequently improving the visual similarity between the original images and the corresponding perturbed ones. However, such a sparsity constraint poses an additional difficulty for the high-dimensional UAA search problem. First, one needs to recognize the pixels which are critical for correctly identifying a majority of the classes. Thereafter, a subset containing a predetermined number of pixels must be selected and efficiently perturbed to obtain the optimal sparse universal perturbation. 

\begin{figure*}[!ht]
    \centering
    \subfigure[\label{fig:model}]{\includegraphics[width=0.75\linewidth]{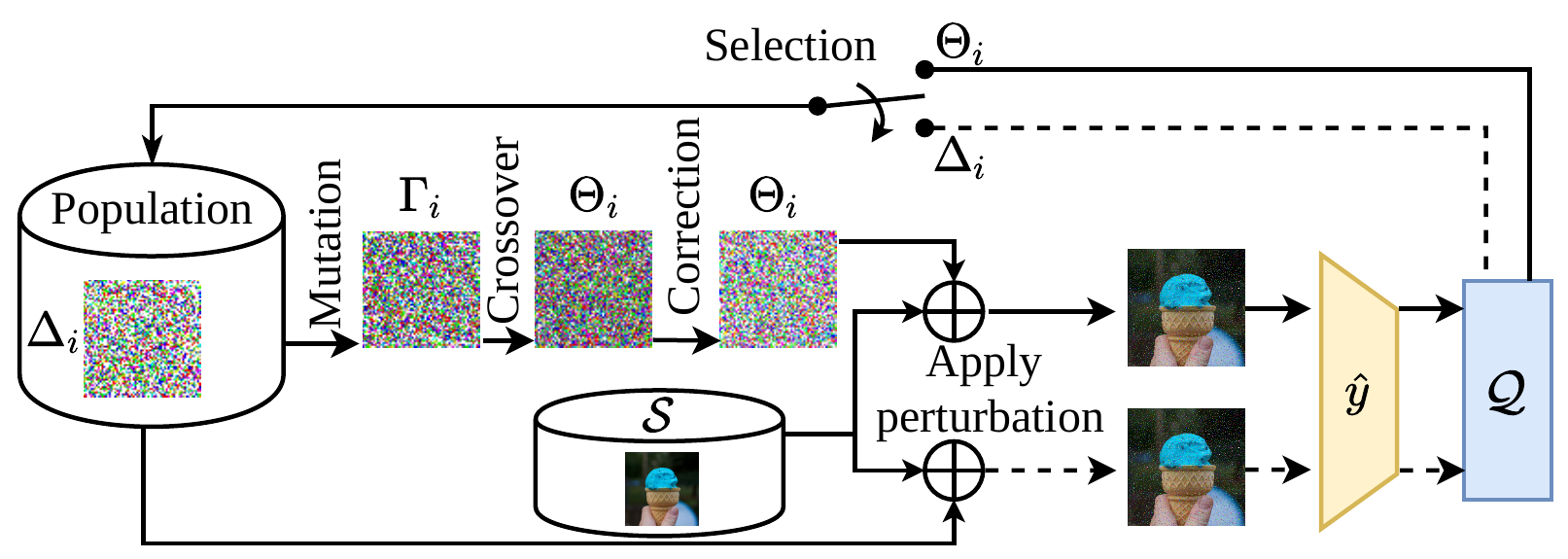}}\\
    \subfigure[\label{fig:encoding}]{\includegraphics[width=0.70\linewidth]{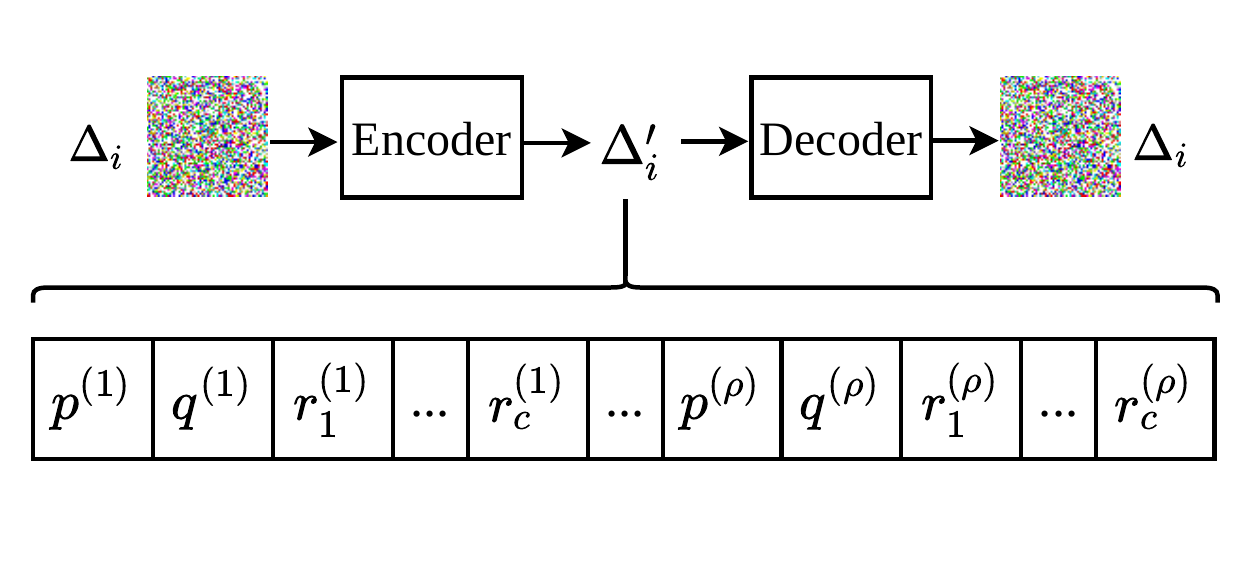}}
    \caption{DEceit model and candidate solution encoding strategy. (a) DEceit attempts to iteratively improve the quality of a population of candidate perturbations (here $\rho=5000$ and $\beta=10$) to find the optimal solution of $\mathcal{Q}$. First a parent perturbation $\Delta_{i}$ for all $i=1, 2, \cdots, n$ is mutated to generate $\Gamma_{i}$. Subsequently, $\Gamma_{i}$ and $\Delta_{i}$ go through crossover to form $\Theta_{i}$. The perturbation $\Theta_{i}$ may not adhere to the constraints of $\mathcal{Q}$, and thus the bound restriction is imposed through correction. Finally, $\mathcal{Q}(\Delta_{i})$ and $\mathcal{Q}(\Theta_{i})$ are both applied to the set $\mathcal{S}$ of images and evaluated. $\Theta_{i}$ replaces $\Delta_{i}$ in the population if found to better optimize $\mathcal{Q}$, i.e. $\mathcal{Q}(\Delta_{i}) \geq \mathcal{Q}(\Theta_{i})$. Note that the perturbations are color inverted for the ease of visualization while their encoding/decoding which might be necessary for the implementation of DEceit are ignored for simplicity. (b) A perturbation $\Delta_{i}$ is encoded as $\Delta_{i}'$ for DEceit. The Encoder performs a one-to-one mapping from perturbation space to a $(c+2)\times\rho$ dimensional encoded space. An encoded perturbation can be considered as $\rho$ groups of $(c+2)$ elements. Given such a group the first two elements $p^{(i)}$ and $q^{(i)}$ denote the position respective to width and height of the $i$-th pixel to be perturbed, where $i=1, 2, \cdots ,\rho$. Further, the last $c$ places $r^{(i)}_{1}, r^{(i)}_{2}, \cdots, r^{(i)}_{c}$ hold the noises for the $c$ channels. Such an encoding can directly handle the sparsity constraint (\protect\ref{eqn:opt2:c2}) by implicitly considering the value of $\rho$ during design. The Decoder performs the reverse mapping to retrieve the original perturbation from its corresponding encoded representation. }
    \label{fig:modelEncode}
\end{figure*}

\subsection{Black-Box Feedback}
The challenge of finding an effective adversarial perturbation further increases in a black-box setting, which restricts utilization of the image-to-class decision mappings expressed through the network architecture and learned parameters. However, such restrictions allow a black-box method to be independent of the CNN under attack. Consequently, perturbation obtained through black-box methods is expected to provide improved robustness and transferability compared to those found by white-box attacks. For a black-box setting, one may choose to maximize the number of images that are predicted differently after applying perturbation, instead of directly maximizing the probability of an alternative labelling. Thus, the optimization problem in (\ref{eqn:opt1}) can be revised for a sparse black-box setting as follows:
\begin{equation}
\label{eqn:opt2}
    \max \mathcal{Q} (\Delta) = \sum_{I \in \mathcal{X}} \mathcal{I}({y}(\psi(I+\Delta)) \neq y(I)), 
\end{equation}
\begin{equation}
    \label{eqn:opt2:c1} \text{s.t.} \; -\beta \leq \Delta_{pqr} \leq \beta, 
\end{equation}
\begin{equation}
    \label{eqn:opt2:c2} \text{and} \; \sum_{p=1}^{w}\sum_{q=1}^{h} \mathcal{I}(\Delta_{pq} \neq 0)=\rho,
\end{equation}
where $\mathcal{I}(\cdot)$ is an indicator function which outputs 1 if the input condition is true and 0 otherwise, $\Delta_{pq}$ denotes a pixel in $\Delta$, and $\rho$ is an user defined parameter controlling the allowable number of pixels which can be perturbed.

\subsection{Differential Evolution--based Optimization}\label{sec:DE}
Due to the black-box setting and the complex nature of problem $\mathcal{Q}$ in (\ref{eqn:opt2}), standard gradient-based optimization cannot be used to obtain an efficient solution. Instead, one may employ EvO techniques in an attempt to find a near-optimal solution after a finite number of generations. An EvO algorithm achieves this by iteratively improving the quality of a population of candidate solutions through exploration (diversifying coarse search in large volume) and exploitation (intensifying detailed search in a small volume) of the search space \cite{opara2019}. Among the various EvO techniques tailored for diverse optimization problems, DE is known to be highly effective for real-valued, global optimization problems like $\mathcal{Q}$ \cite{das2016recent}. Due to the large-scale and high-dimensional nature of $\mathcal{Q}$, we specifically incorporate some modifications to equip DE for the task, resulting in the DEceit algorithm.

\subsubsection{Algorithm Overview:}
Before proceeding to a detailed discussion, we first provide a schematic workflow of DEceit in Figure \ref{fig:model} to highlight the key steps involved in the process. Given the problem $\mathcal{Q}$, the DEceit algorithm starts by randomly initializing a population $N=\{\Delta_{1}, \Delta_{2}, \cdots \Delta_{n}\}$ of $n$ candidate perturbations. However, simply representing a perturbation as a natural tensor of the same size of the original image will not be effective as the different evolutionary operations may end up violating the sparsity constraint (\ref{eqn:opt2:c2}). Thus, during the DE operations, instead of the original perturbation tensor $\Delta_{i}$ we consider an encoded vector representation $\Delta'_{i}$ that adheres to (\ref{eqn:opt2:c2}) by design, having a dimension of $(c+2)\times\rho$ as detailed in Figure \ref{fig:encoding}. Hereafter, we denote an encoded perturbation using a dashed notation of their tensor counterpart. Subsequently, every iteration of DEceit consists of the following four steps: First, a trial solution $\Gamma_{i}'$ corresponding to each of the current solutions $\Delta_{i}'$ is created as a linear combination of three other candidate solutions using one of the following mutation schemes as explained in the subsequent section discussing mutation switching strategy:
\begin{equation}
\text{DE/rand/1: } \Gamma_{i}' = \Delta_{u1}' + F (\Delta_{u2}' - \Delta_{u3}'), \text{ or} \label{eqn:mutrand}
\end{equation}
\begin{equation}
\text{DE/best/1: } \Gamma_{i}' = \Delta_{b}'  + F (\Delta_{u2}' - \Delta_{u3}'), \label{eqn:mutbest}
\end{equation}
where $F$ is the scale-factor, $\Delta_b'$ is the encoded representation of the best perturbation in $N$, and $\Delta_{u1}'$, $\Delta_{u2}'$, $\Delta_{u3}'$ are selected randomly from $N$ without replacement and differ from the current solution $\Delta_i'$. Following mutation, $\Gamma_{i}'$ goes through an arithmetic crossover operation as shown in (\ref{eqn:cross}) to generate the final offspring solution $\Theta_{i}'$ as average of $\Gamma_{i}'$ and the corresponding candidate solution $\Delta_{i}'$:
\begin{equation}
    \Theta_{i}' = 0.5 \Delta_{i}' + 0.5 \Gamma_{i}'. \label{eqn:cross}
\end{equation}
The generated offspring vector $\Theta_{i}'$ is subsequently corrected (as explained during the discussion on bound correction) to adhere to constraint (\ref{eqn:opt2:c1}) and decoded to retrieve the perturbation $\Theta_{i}$. Finally in the selection stage, $\Theta_{i}$ is compared with the corresponding decoded candidate solution $\Delta_{i}$ so as to greedily retain the encoded form of the better among the two in the population $N$:
\begin{equation}\label{eq:selec}
    \Delta_{i}' = \begin{cases}
                 \Theta_{i}' & \text{if}\ \mathcal{Q}(\Theta_{i}) \geq \mathcal{Q}(\Delta_{i}), \\
                 \Delta_{i}' & \text{otherwise}.
                 \end{cases}
\end{equation}

\subsubsection{Mutation Switching Strategy:}
Only exploratory search will result in delayed or no convergence whereas an only exploitative search might push the population towards a local optimum. Striking a proper balance between these two extremes is critical for high-dimensional search scenarios. The DEceit search mechanism is controlled together by the scale factor $F$ and the chosen mutation strategy. For every candidate solution in the population, we randomly switch the value of $F$ between only a couple of choices, namely 0.5 and 2 with uniform probability. When $F$ attains the higher value of 2, the greater weighting of the difference $(\Delta_{u2}' - \Delta_{u3}')$ helps to the formation of a trial solution that is distant from the base solution ($\Delta_{u1}'$ or $\Delta_b'$ depending on the chosen mutation strategy). On the other hand, when $F$ attains the lower value of $0.5$, the trial solution lies close to the current base member ($\Delta_{u1}'$ or $\Delta_{b}'$), leading to exploitation of already identified regions of the search space. Thus. these choices of $F$ selected in a uniformly random manner can help stagnant solutions out of local minima. 

While the `DE/rand/1' mutation strategy encourages exploration or exploitation based on the chosen $F$, `DE/best/1' draws the population close to the current best solution. DEceit chooses either of the two mutation strategies for every candidate solution, in every generation, based on the success history in the last generation. This means that the same mutation strategy as the last generation is retained only if it had resulted in a final offspring superior to the corresponding candidate solution. Thus, the scale factor and mutation strategy together decide the trade-off between the exploratory and exploitative searching capability of DEceit.

\subsubsection{Bound Corrections:}
While the encoding implicitly handles the sparsity constraint (\ref{eqn:opt2:c2}), the algebraic operations in mutation and crossover may still result in an invalid offspring. Considering the case for the $i$-th pixel to be perturbed where $i=1, 2, \cdots, \rho$, a constraint violation may happen in the following three ways, all of which can be easily rectified. First, if any of the location specifiers $p^{(i)}$ and $q^{(i)}$ is not an integer, then they can be set to their ceiling values. Second, if any of $p^{(i)}$ and $q^{(i)}$ do no respectively lie in the range of $[1, w]$ and $[1, h]$, then they can be reinitialized in their corresponding range. Third, if any of $r^{(i)}_{1}, r^{(i)}_{2}, \cdots, r^{(i)}_{c}$ fails to satisfy the bound constraint (\ref{eqn:opt2:c1}), then the concerned component can be rectified by reinitializing with a real number between $-\beta$ and $\beta$. We detail the complete pseudo-code in the following Algorithm 1. 

\subsubsection{Complexity of DEceit:}
A common concern regarding the EvO techniques is that they are likely to increase the time complexity. To evaluate if the EvO based nature of DEceit adversely affects an economic performance we present its asymptotic time complexity in Theorem \ref{timeCompTheo}. 

\begin{theorem}
\label{timeCompTheo}
If a perturbation is allowed to perturb only $\rho$ pixels in an image then the asymptotic time complexity of DEceit can be expressed as $O(\rho)$.
\end{theorem}
\begin{proof}
As we are attacking a deep learning network that is known to effectively exploit the opportunities of parallelism our implementation of DEceit should also follow the same philosophy. In DEceit, during initialization, we represent each of the initial $n$ candidate perturbations through a $O(\rho)$-dimensional encoded vector where $\rho$ is the number of pixels to be perturbed. In each iteration, DEceit takes $n$ optimization steps corresponding to each of the $n$ candidate solutions. In each step, a candidate solution goes through a mutation, crossover, bound correction, decoding, and selection stages. Mutation and crossover operations are simply linear combinations of $O(\rho)$-dimensional vectors thus have a time complexity of $O(\rho)$. Bound correction requires validating each dimension of a candidate solution and decide on necessary adjustments. Evidently, this can be performed in $O(\rho)$ time as well. If a Graphics Processing Unit (GPU) is used to exploit the opportunities of parallelism then mutation, crossover, and bound correction can all be performed in $O(1)$ time as the dimensions of the vector are independent. Now, $O(\rho)$ time will be required during the decoding as there the task is to scan the candidate vector to retrieve each of the $\rho$ pixel positions in the perturbation matrix and accordingly set the corresponding channel values. During selection, the candidate perturbation is first added with the image of the same dimension which takes $O(1)$ time using parallel processing. Further, to evaluate the performance of a candidate perturbation the class label of the perturbed image is predicted by the network to be attacked which takes $O(\lambda)$ time. However, in a given problem, the prediction time of the attacked network can be considered a constant. Therefore, each of the $n$ candidate solutions takes a total of $O(\rho)$ time to evolve and subsequently evaluated. Note that, the switching of $F$ and mutation scheme can both be performed in constant times as well. Thus, in $t$ iterations the total time complexity of DEceit is reduced to $O(tn\rho)$. Again, $t$ and $n$ are traditionally considered as constant \cite{biswas2018parameter} while the product $tn$ is usually small compared to a $\rho$ which can be as large as $whc$, hence $O(nt\rho) \approx O(\rho)$ completing the proof.
\end{proof}

From Theorem \ref{timeCompTheo} it is evident that the asymptotic time complexity of DEceit is linearly dependent on the number of perturbed pixels. Thus, DEceit is not only capable of providing a fairly economic solution for the adversarial perturbation problem but can also effectively utilize the scope of sparsity.   

\begin{algorithm}[!ht]
  \caption{DEceit}
  \label{alg:deceit}
  \footnotesize
  {\bfseries Input:} $\mathcal{X}$: Set of images for UAA/ singleton for IDA. $\rho$: Number of allowable pixels for perturbation (if sparsity is not required then set $\rho=wh$). $t$: Maximum allowable epochs, $n$: Population size. \\
  {\bfseries Output:} An adversarial perturbation $\Delta$.
  \let \oldnoalign \noalign
  \let \noalign \relax
  \midrule
  \let \noalign \oldnoalign
  \begin{algorithmic}[1]
  \STATE Randomly initialize $n$ perturbations $\Delta_{1}, \Delta_{2}, \cdots, \Delta_{n}$ and corresponding $s_{i}$ where $\Delta_{i} \in \mathbb{R}^{w \times h \times c}$ follows constraints (\ref{eqn:opt2:c1}-\ref{eqn:opt2:c2}) and $s_{i} \in \{0, 1\}, \forall i \in \llbracket 1, n \rrbracket$. Form population $N=\{\Delta_{1}', \Delta_{2}', \cdots, \Delta_{n}'\}$ by encoding $\Delta_{i}$s and calculate initial fitness $e_{i}=\mathcal{Q}(\Delta_{i}), \forall i \in \llbracket 1, n \rrbracket\}$.
  \FOR{$t$ iterations} 
         \FOR{$i \in \{1, 2, \cdots, n\}$}
            \STATE Select a random scale-factor $F \in \{0.5, 2\}$.
            \STATE If $s_{i}=0$ then generate mutated solution $\Gamma_{i}'$ as per (6). Otherwise find $\Gamma_{i}'$ using (7).
            \STATE Perform crossover to generate perturbation $\Theta_{i}'$ by (8). 
            \STATE Perform bound corrections on $\Theta_{i}'$ and retrieve $\Theta_{i}$.
            \STATE Perform selection as per (9). If $\Delta_{i}$ is selected over $\Theta_{i}$ then set $s_{i}=1-s_{i}$, else set $\Delta_{i}'=\Theta_{i}'$ and $e_{i}=\mathcal{Q}(\Theta_{i})$.
        \ENDFOR
    \ENDFOR
    \STATE Return the best performing decoded perturbation $\Delta_{b}$.
\end{algorithmic}
\end{algorithm}

\section{Experimental Evaluation}\label{sec:exp}
In this section, we first discuss the experimental protocol and the implementation in detail. Subsequently, we illustrate the development of DEceit through an ablation study. Finally, we present the performance of DEceit in comparison to the state-of-the-art methods.

\subsection{Experimental Protocol}
All of our experiments are performed on ImageNet 2012 dataset \cite{ILSVRC15}. For performing successful adversarial attacks during the ablation study and comparative evaluation with other UAA techniques we first train DEceit on the entire ImageNet training set. We then test the performance of the perturbation produced by the trained DEceit on 50000 ImageNet 2012 validation samples spread across 1000 classes. Depending on the contending algorithm, the state-of-the-art CNN classifiers attacked by DEceit are VGG16 \cite{simonyan2014very}, GoogleNet \cite{szegedy2015going}, Inception V3 \cite{szegedy2016rethinking}, and ResNet 50 \cite{he2016deep}.  

To quantitatively evaluate the performance of DEceit we have used the conventional Fooling Rate index, while the visual similarity of the perturbed image with the corresponding original one is measured by Peak Signal to Noise Ratio (PSNR) \cite{hore2010image}. To attenuate any bias which may originate from the randomness implicit to DEceit, we report the median result among 11 independent runs in all our experiments. The number of population members and the number of iterations of DEceit are both set to 50.

In case of UAA during selection step the expression \ref{eq:selec} calculates the loss over the entire training set. If the training set is large enough then such an approach may severely limit the computational feasibility of DEceit. However, DEceit employs an asynchronous version of DE\footnote{DE can be either synchronous or asynchronous in nature. Similar to batch gradient descent, in synchronous DE a selected offspring has to wait for the iteration to end before it can replace its parent in the population. On the other hand, in asynchronous DE, analogous to stochastic gradient descent, the offspring is immediately included in the population if it is selected.} that allows us to only consider a subset of the training set during selection. Evidently, the cardinality of the subset should not be too small as that may reduce diversity while it cannot also be large enough to impact the computational cost. We find that calculating loss over a subset (randomly selected without replacement from the training set with uniform probability) of 20000 images works consistently well on average and thus all our experiments are performed using that setting.  

\begin{figure}[!ht]
    \centering
    \includegraphics[width=0.8\textwidth]{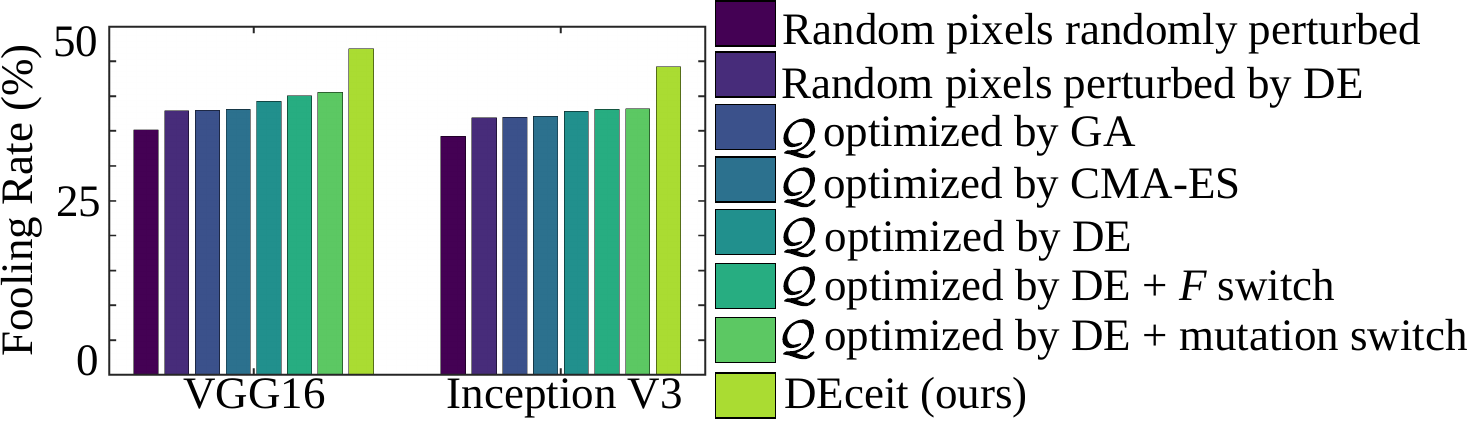}
    \caption{Ablation study of DEceit on performing UAA. For all cases $\beta$ is set to 128 while $\rho$ is set to 10 except the cases of random pixel perturbations (either randomly or by DE). In cases of the two exceptions we allow higher $\rho=200$ to compensate for an unoptimized perturbation.} 
    \label{fig:ablation}
\end{figure}

\subsection{Ablation Study}\label{sec:ablation}

We start by justifying the motivation behind DEceit through an ablation study detailed in Figure \ref{fig:ablation}. We can observe that compared to a benchmark of absolute random perturbations, better UAA can be performed if only the distortions are allowed to be optimized by the EvO algorithm such as DE at random pre-determined pixel positions. If we allow both the pixel locations and the distortion to be optimized by EvO algorithms by maximizing $\mathcal{Q}$, then an even higher Fooling Rate can be achieved despite the number of perturbed pixels being greatly reduced. However, compared to other widely popular EvO algorithms such as the GA variant used in \cite{alzanot2019genattack} and CMA-ES \cite{hansen2006cma}, DE achieves a better Fooling Rate validating its applicability to the problem of adversarial attack generation. Note that, for elitist GA the population size and iteration number, both are chosen as $50$. For CMA-ES, a regular optimizer without any modifications such as restarting is used following the recommended practice
\cite{hansen2006cma}. Moreover, in the above experiments where a regular DE with DE/rand/1 scheme is used, the scale factor, crossover probability, number of population members, and number of generations are respectively set to 0.8, 0.5, 50, and 50. However, as mentioned earlier, vanilla DE is not tailored to handle a large-scale, high-dimensional optimization problem. Thus, we validate the efficacy of the proposed modifications namely switching $F$ and mutation schemes. Even though switching $F$ or the mutation schemes alone is capable of achieving better performance, the improvement is only marginal. Thus, we finally replace DE with DEceit which employs both $F$ and mutation schemes switching strategies. It is evident from Figure \ref{fig:ablation} that DEceit can indeed offer a considerable performance improvement.

\begin{table*}[!ht]
    \centering
    \caption{Comparison of DEceit with non-targeted UAA methods on ImageNet 2012 validation set.}
    \label{tab:uapRes}
    \vspace{5pt}
    \begin{threeparttable}
    \scriptsize
    \begin{tabular}{ccc|ccc|ccc} \toprule
        Method & $\rho$ & $\beta$ & \multicolumn{3}{c|}{Fooling Rate (\%)} & \multicolumn{3}{c}{PSNR (dB)} \\
        & & & VGG16 & Inception V3 & GoogleNet & VGG16 & Inception V3 & GoogleNet \\ \midrule
        UAP    & 50176 & 10 & 78.37  & -  &    77.74    &    29.25    &   - &28.45     \\
        FFF & 50176    & 10 &    47.25 & -  &    56.59    &    27.12    &    - & 24.56     \\
        NAG & 50176 & 10 & 77.57 & \textbf{90.37} & - & - & - & -  \\
        GAP & 50176    & 10 & 83.70 & 82.70 & - & - & - & - \\
        \midrule
        \multirow{6}{*}{\shortstack{ DEceit \\ (Ours)} } & 10 & 128 &    46.79 & 44.23    & 44.44 &   \textbf{65.12}    &    \textbf{65.01} & \textbf{68.99}  \\
        & 100 & 30    & 56.22	& 52.86 & 54.55    &    56.76    &    57.01  & 58.69 \\
        & 1000    & 20 &    62.05	& 53.11  & 57.69  &    48.25    &    48.22 & 49.58    \\
        & 5000 & 10 &    65.66	& 59.55 & 61.56   &    44.16    &    45.01  & 46.89  \\
        & 10000    & 10 &    68.95    &    63.25 & 66.89    &    38.96    &    37.57  & 39.44  \\
        & 50176 & 10 & \textbf{83.96}    &    83.03 &  \textbf{84.96}  &    33.14    &    32.99   & 33.25 \\ \bottomrule
    \end{tabular}
    \begin{tablenotes}
    \item ''-" indicates that the pre-trained perturbations are not available for computing Fooling Rate \& PSNR. 
    \item The best results are boldfaced. 
    \end{tablenotes}
    \end{threeparttable}
\end{table*}

\begin{figure}[!ht]
    \centering
    \includegraphics[width=0.7\textwidth]{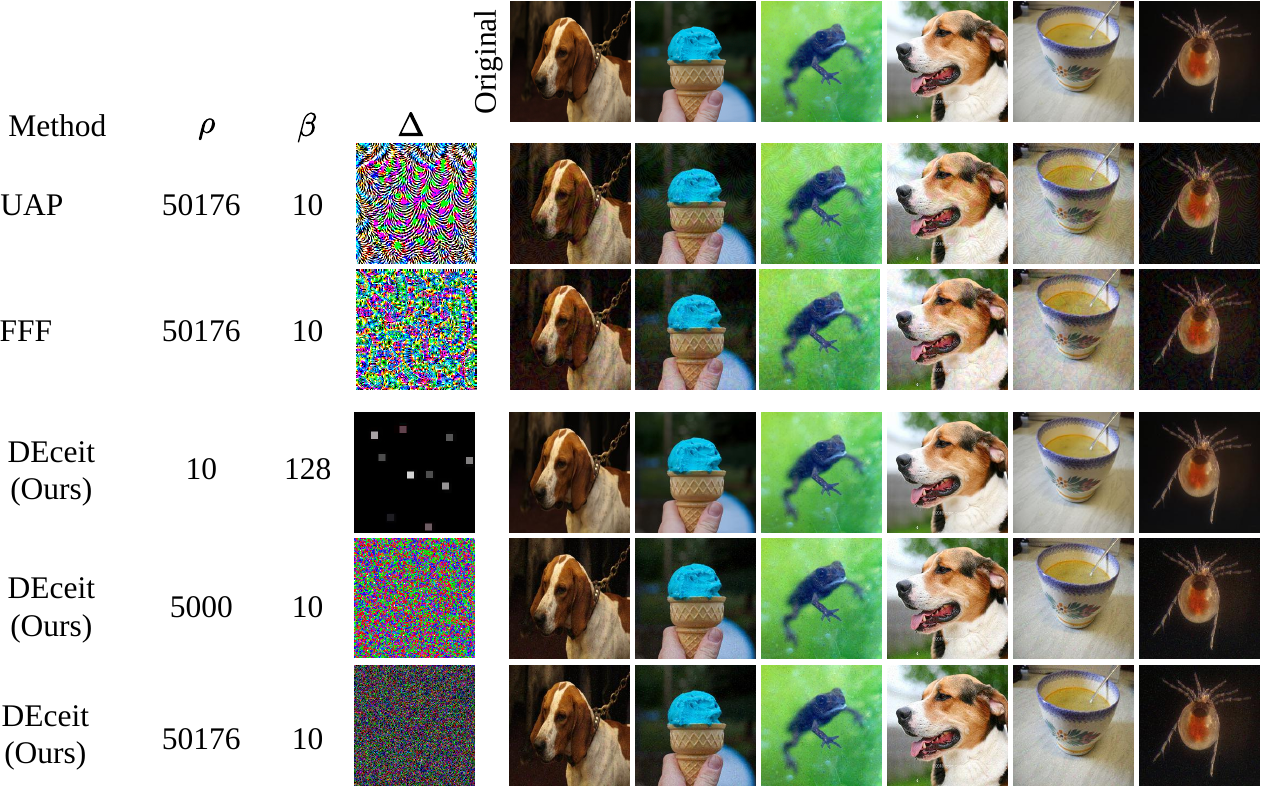}
    \caption{To qualitatively assess the performance of DEceit we demonstrate 6 successful attacks on VGG16 using the universal perturbations generated by UAP, FFF, and DEceit variants. We can see that compared to white-box methods, DEceit can generate a successful adversarial example with minimal distortion. The effect of perturbation (values of the perturbed pixels are propagated to their neighbors for ease of visualization) is almost imperceptible when $\rho=10$. Even when $\rho=50176$ the perturbed images generated by DEceit are of better visual quality than UAP and FFF.} 
    \label{fig:visual}
\end{figure}

\begin{figure}[!ht]
    \centering
    \subfigure[\label{fig:uapvggGnet}]{\includegraphics[width=0.48\linewidth]{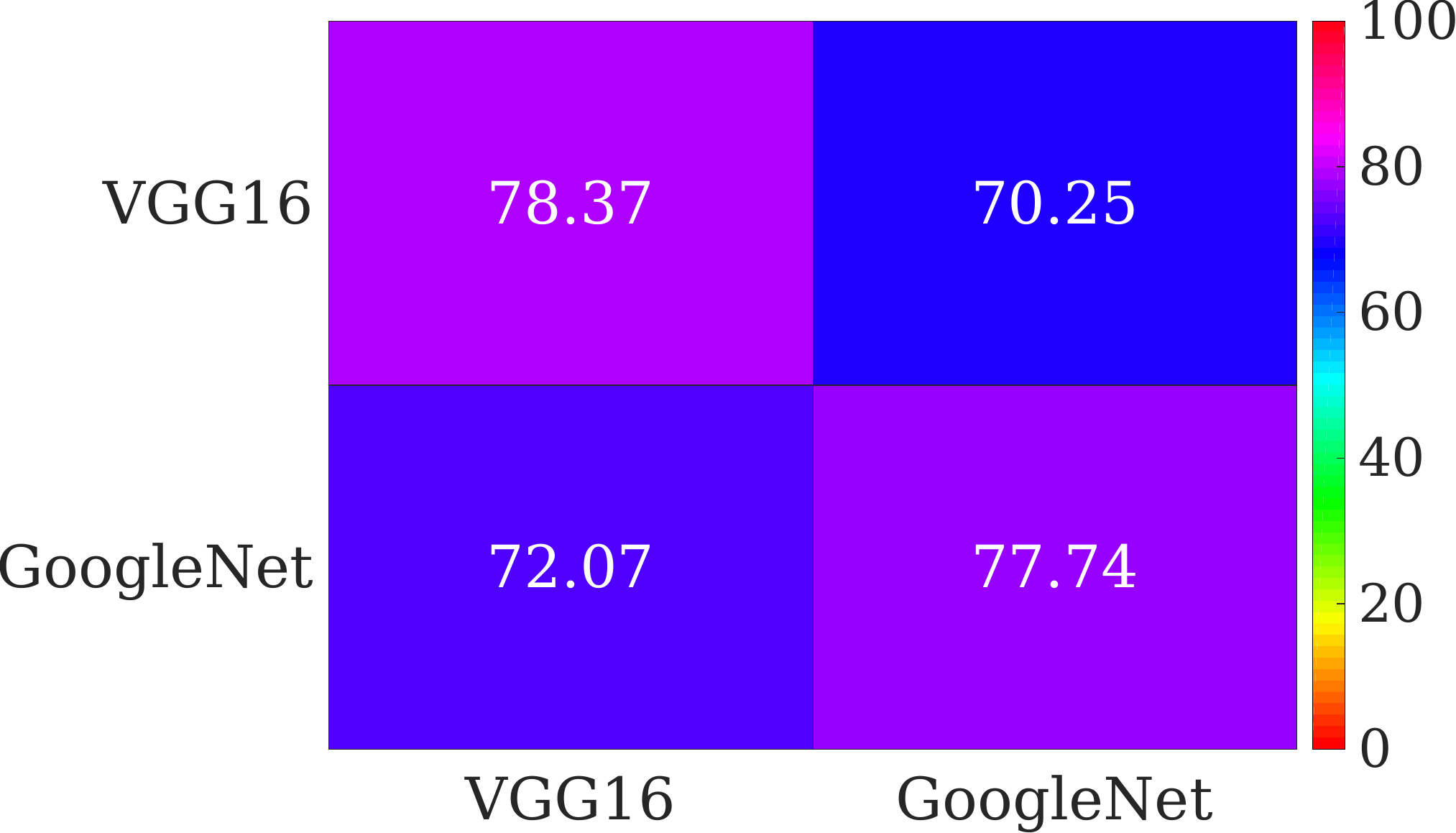}} \hspace{5pt}
    \subfigure[\label{fig:fffvggGnet}]{\includegraphics[width=0.48\linewidth]{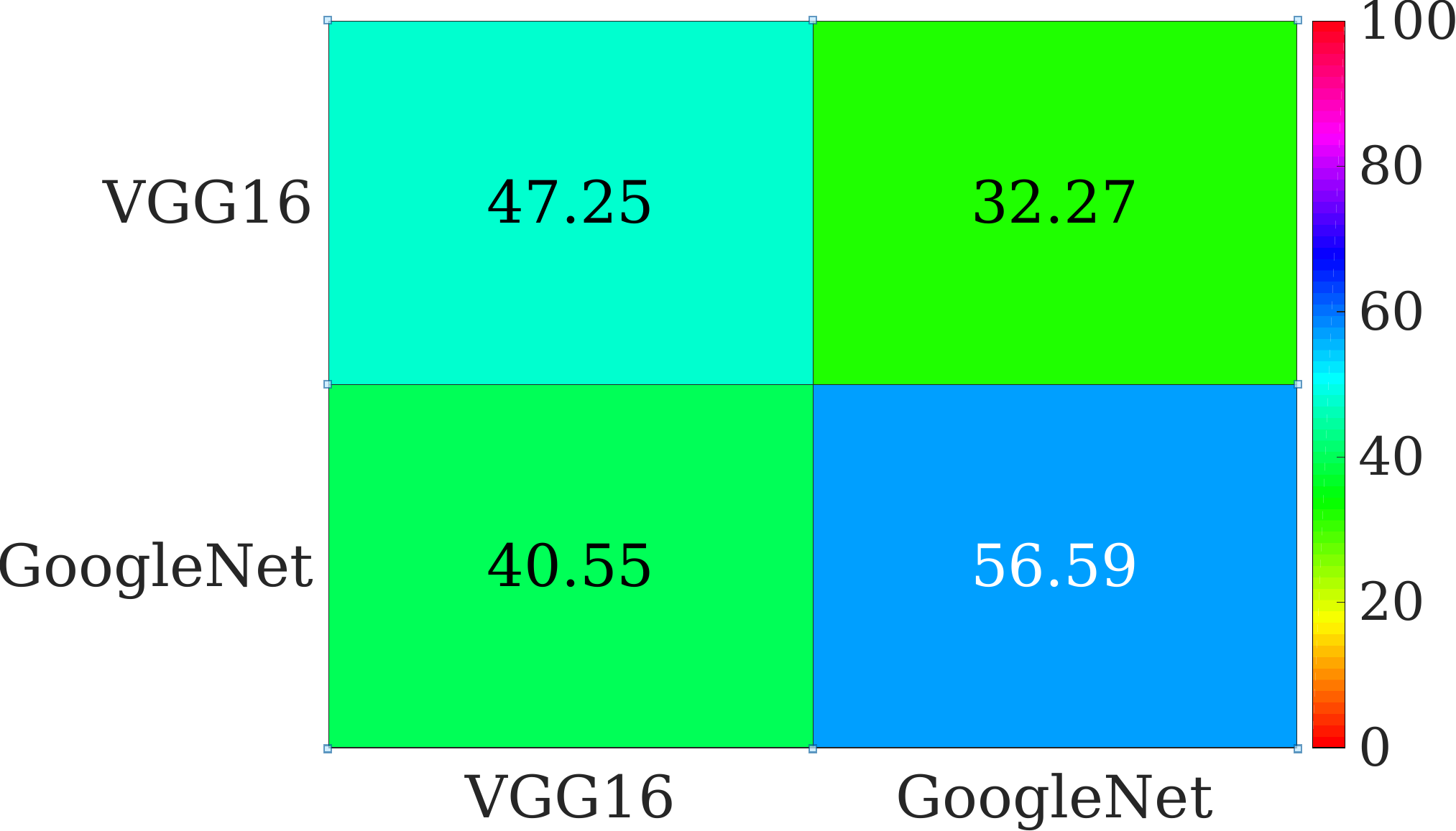}} 
    \subfigure[\label{fig:deceit10128vgggnet}]{\includegraphics[width=0.48\linewidth]{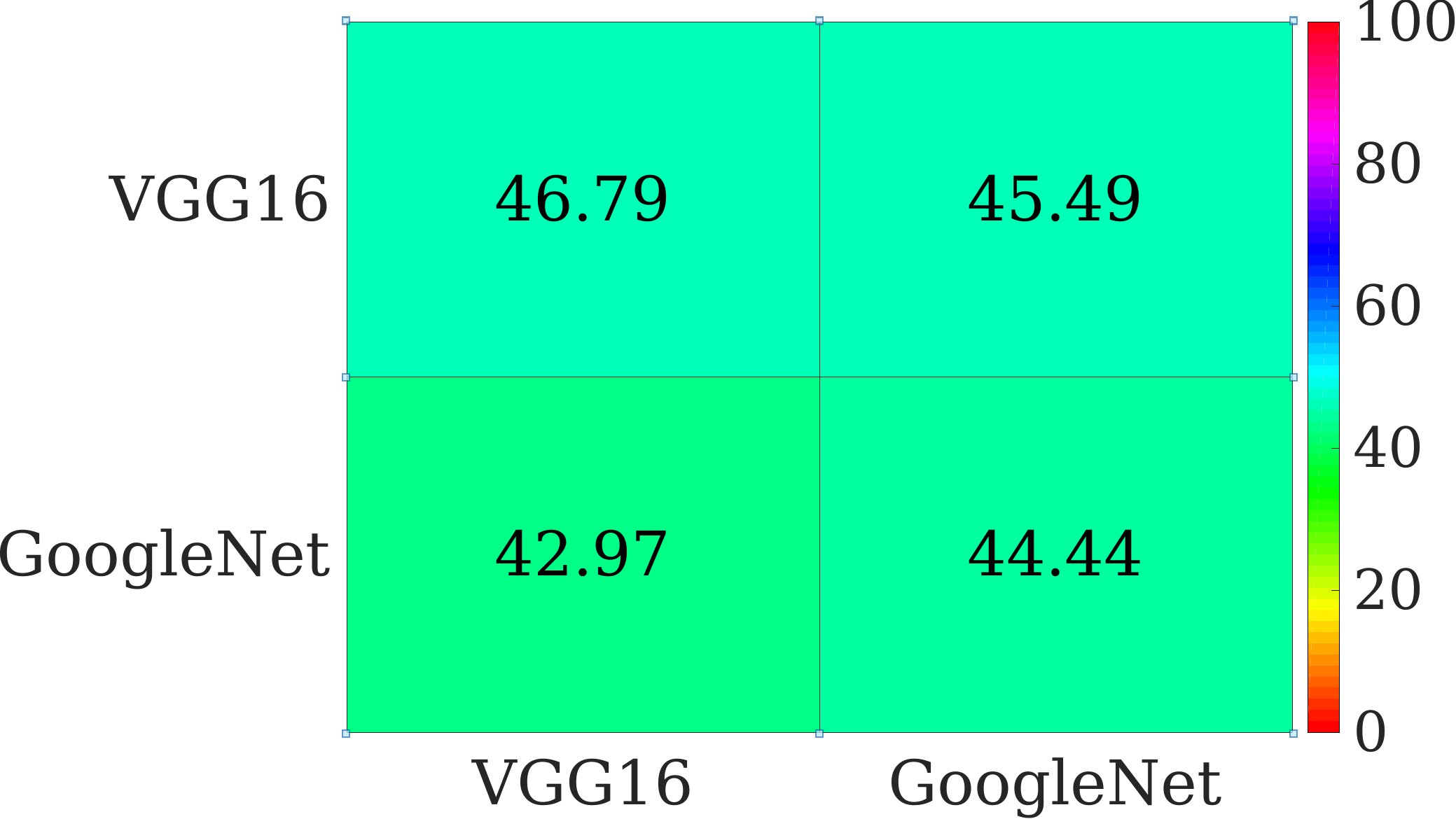}} \hspace{5pt}
    \subfigure[\label{fig:deceit5k10vgggnet}]{\includegraphics[width=0.48\linewidth]{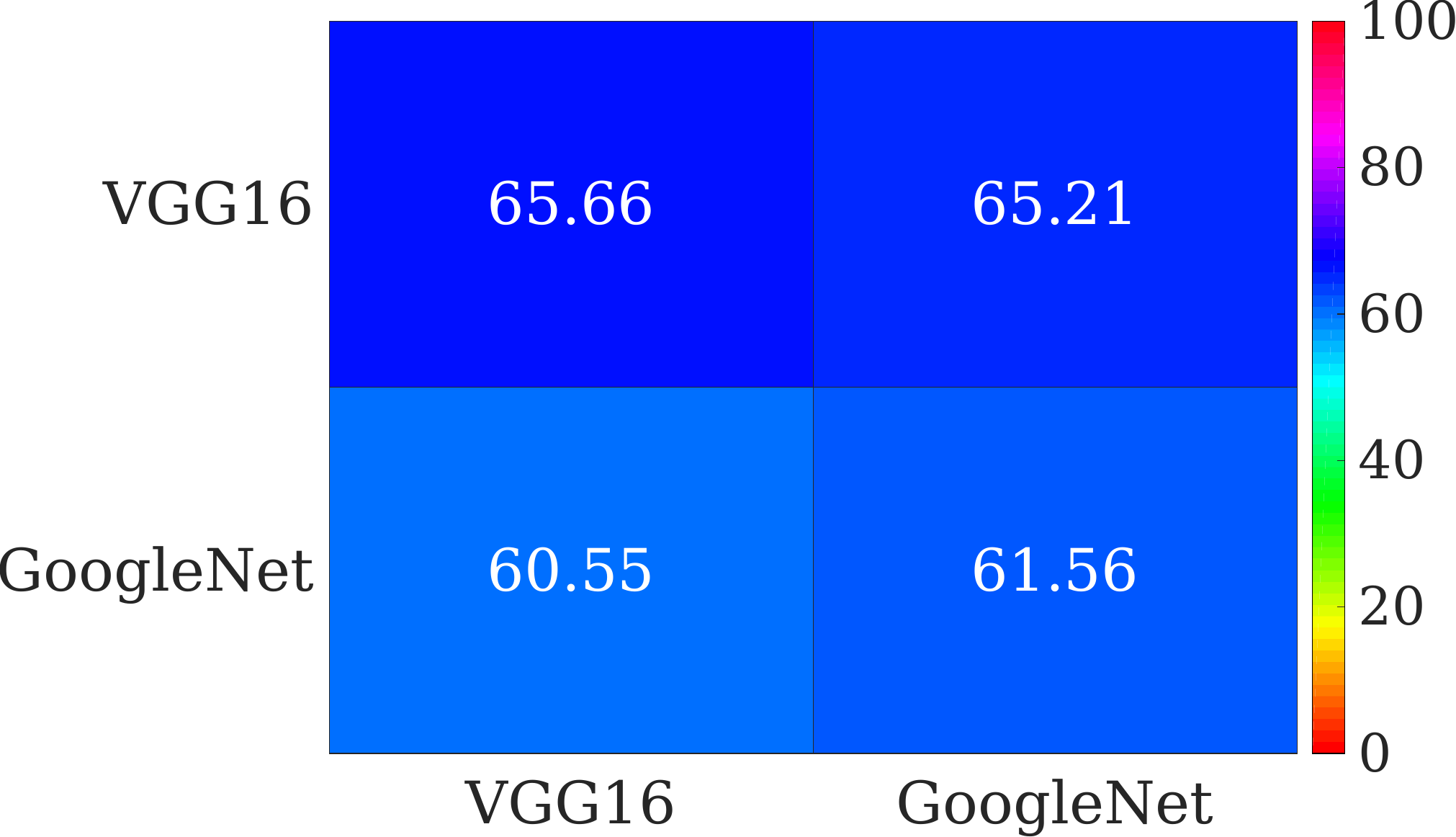}} 
    \subfigure[\label{fig:deceitall10vgggnet}]{\includegraphics[width=0.48\linewidth]{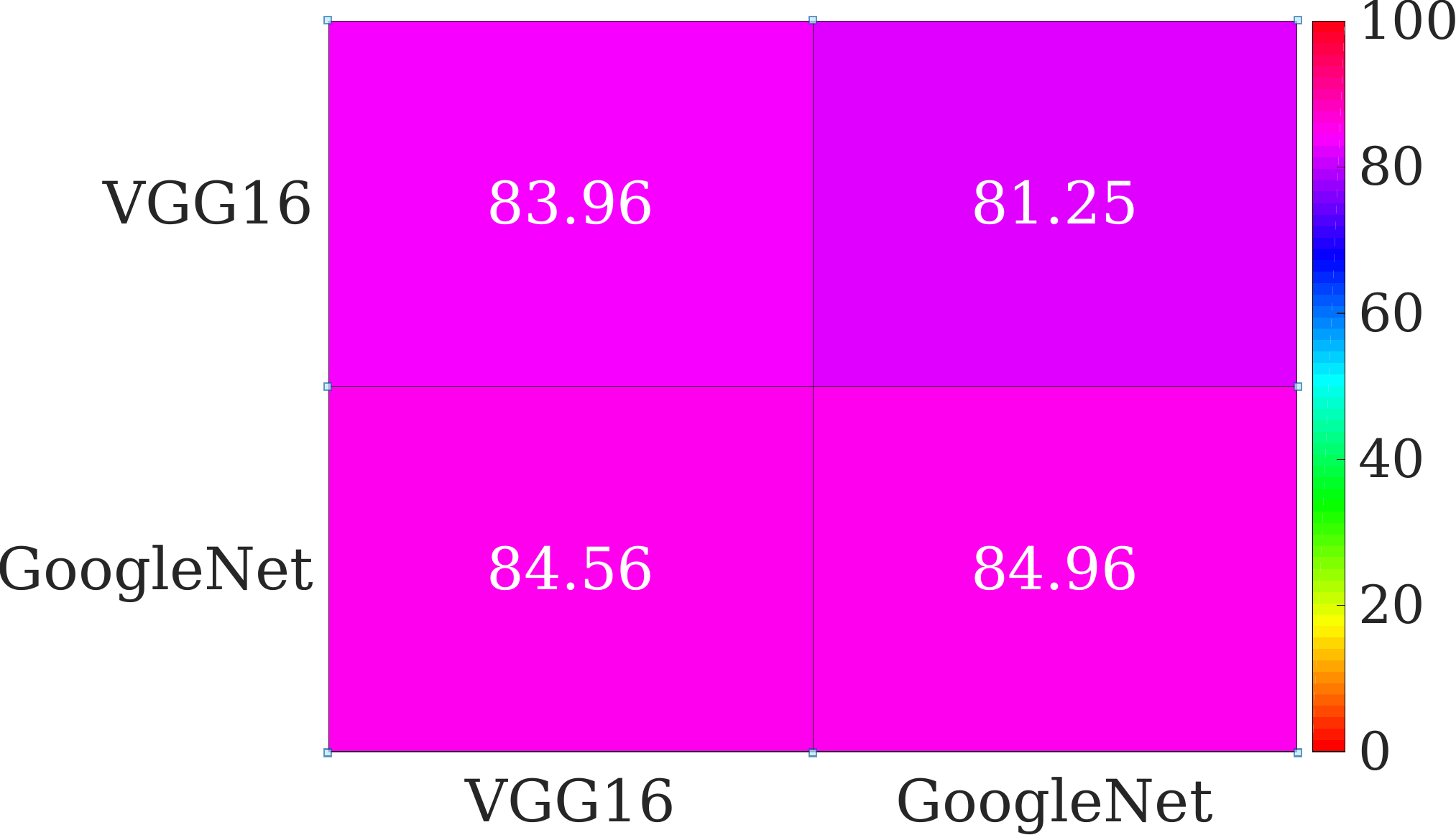}} 
    \caption{Transferability of universal perturbation obtained by DEceit and its competitors on VGG16, GoogleNet. The DEceit variants are named as DEceit-$\rho$-$\beta$. The entry in the $i$-row and the $j$-th column describes the Fooling Rate achieved by applying the universal perturbation optimized for the network in the $i$-th row to the network in the $j$-th column. Transferability comparison of universal perturbation generated by  (a) UAP (b)  FFF (c) DEceit-10-128 (d) DEceit-5000-10 (e) DEceit-50176-10.}
    \label{fig:transferability_raw_vgg_gnet}
\end{figure}

\begin{figure}[!ht]
    \centering
    \subfigure[\label{fig:nagvggIv3}]{\includegraphics[width=0.48\linewidth]{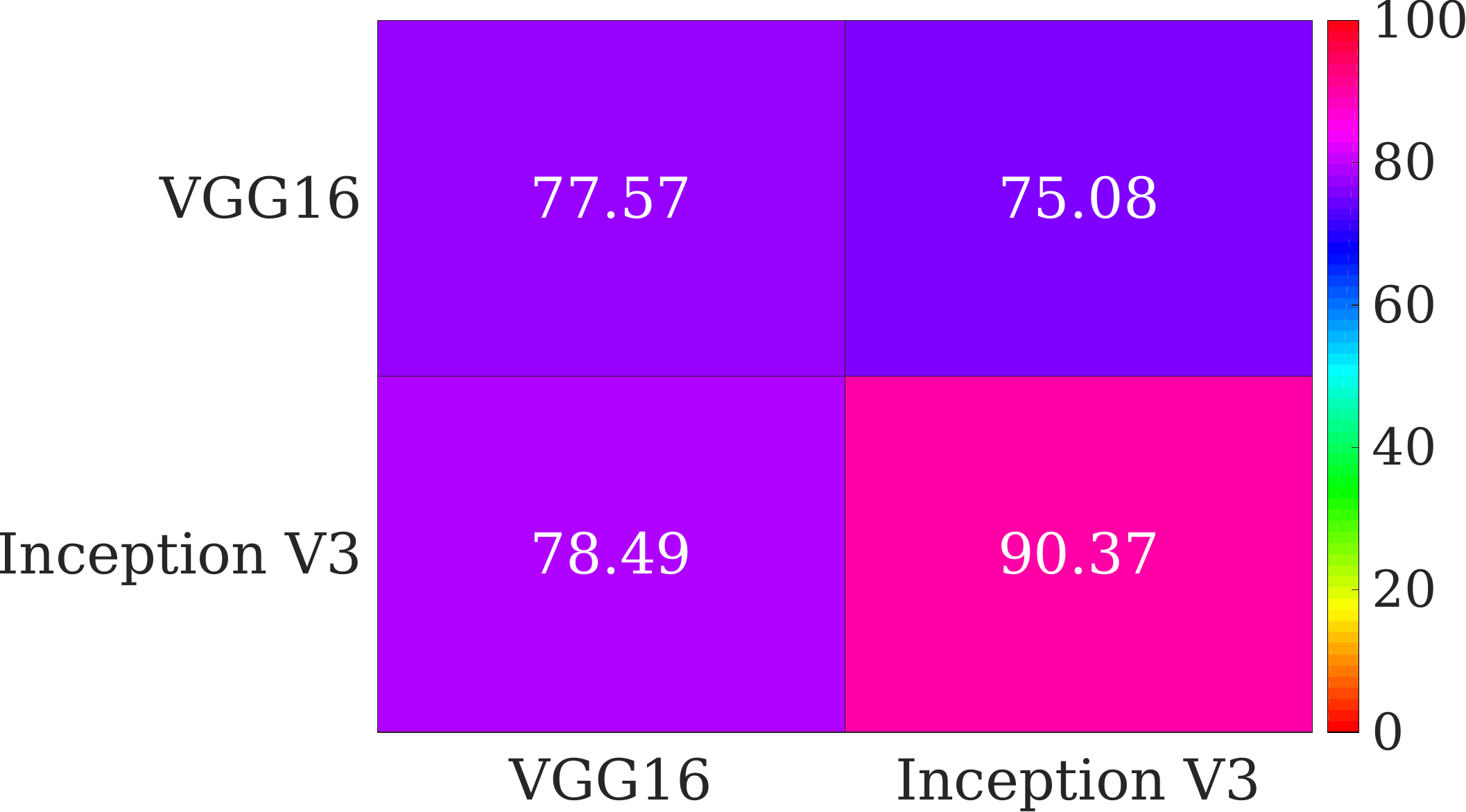}} \hspace{5pt}
    \subfigure[\label{fig:deceit10128vggincv3}]{\includegraphics[width=0.48\linewidth]{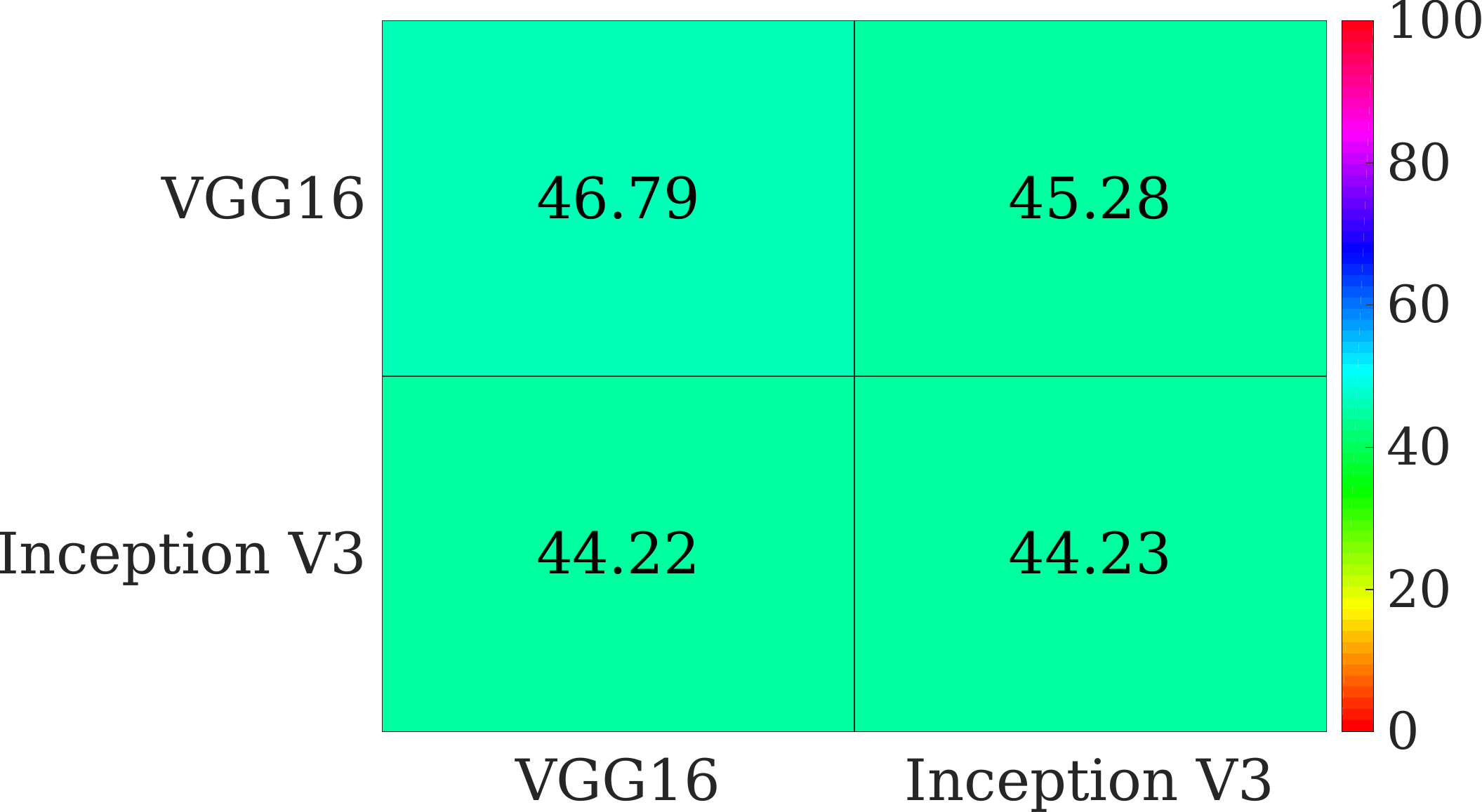}} 
    \subfigure[\label{fig:deceit5k10vggincv3}]{\includegraphics[width=0.48\linewidth]{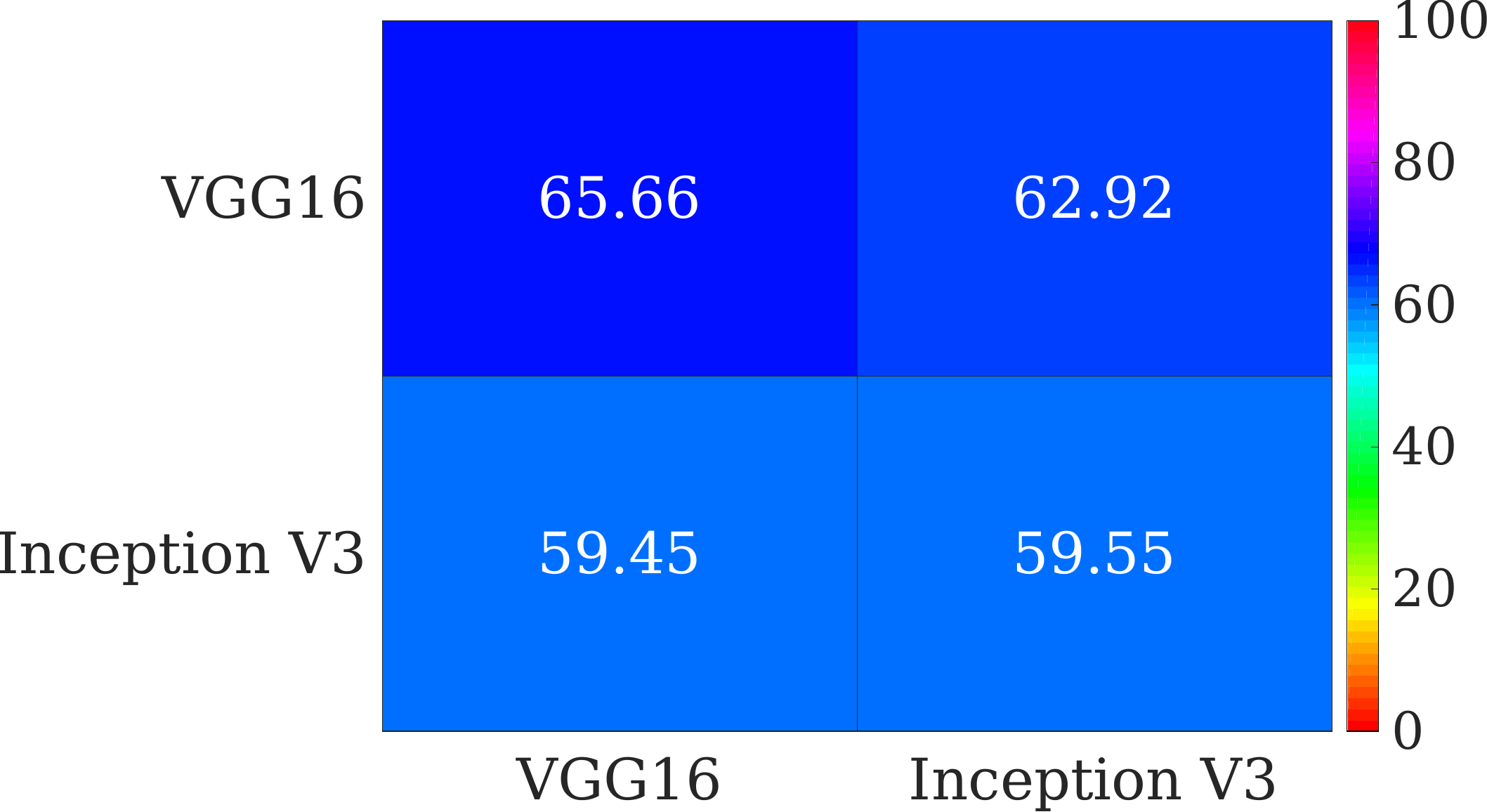}} \hspace{5pt}
    \subfigure[\label{fig:deceitall10vggincv3}]{\includegraphics[width=0.48\linewidth]{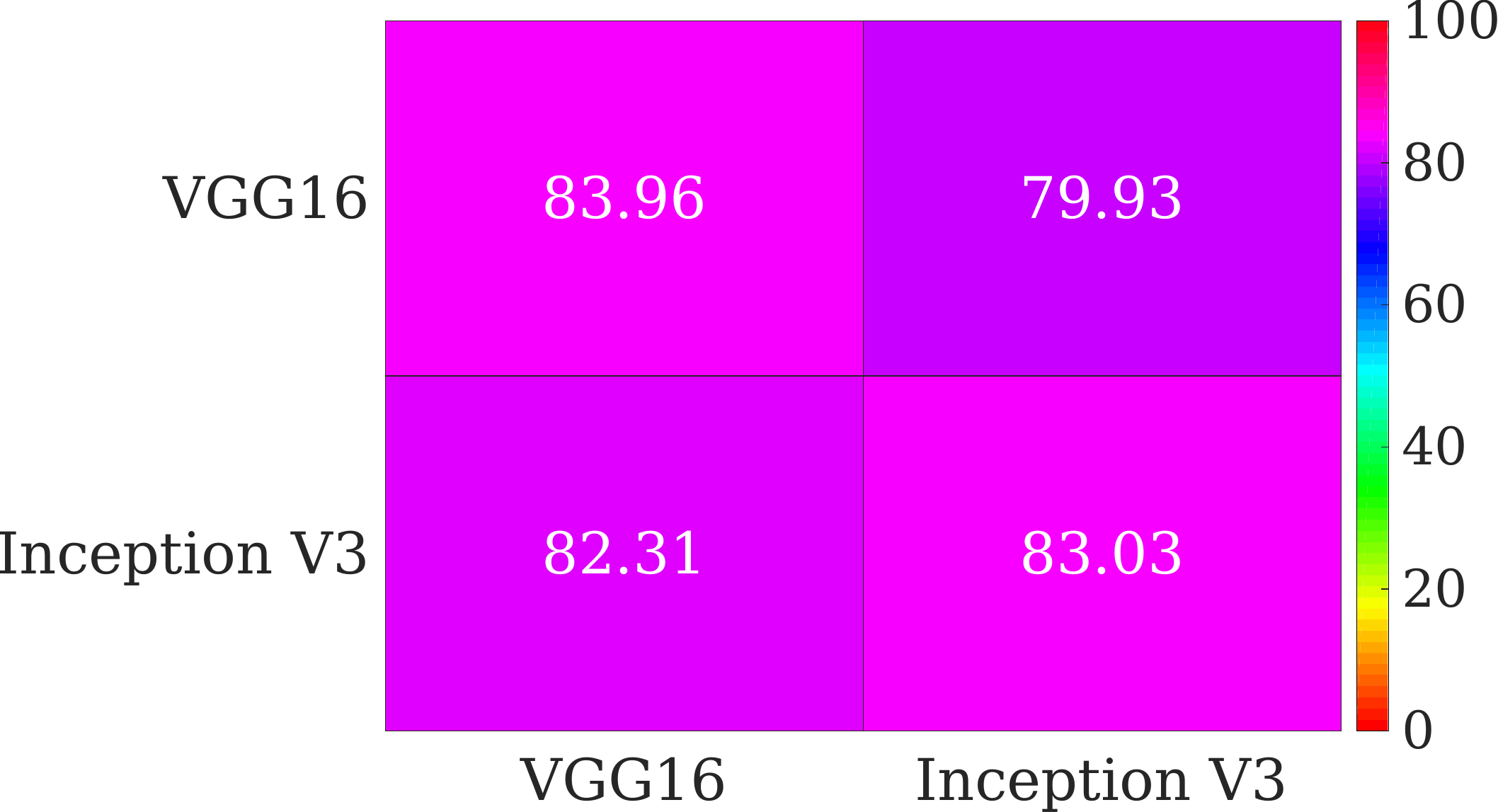}} 
    \caption{Transferability of universal perturbation obtained by DEceit and NAG on VGG16 and InceptionV3. The DEceit variants are named as DEceit-$\rho$-$\beta$. The entry in the $i$-row and the $j$-th column describes the Fooling Rate achieved by applying the universal perturbation optimized for the network in the $i$-th row to the network in the $j$-th column. Transferability comparison of universal perturbation generated by  (a) NAG  (b) DEceit-10-128 (c) DEceit-5000-10 (d) DEceit-50176-10.}
    \label{fig:transferability_raw_vgg_incv3}
\end{figure}

\begin{figure}[!ht]
    \centering
    \subfigure[\label{fig:vggGnet}]{\includegraphics[width=0.35\linewidth]{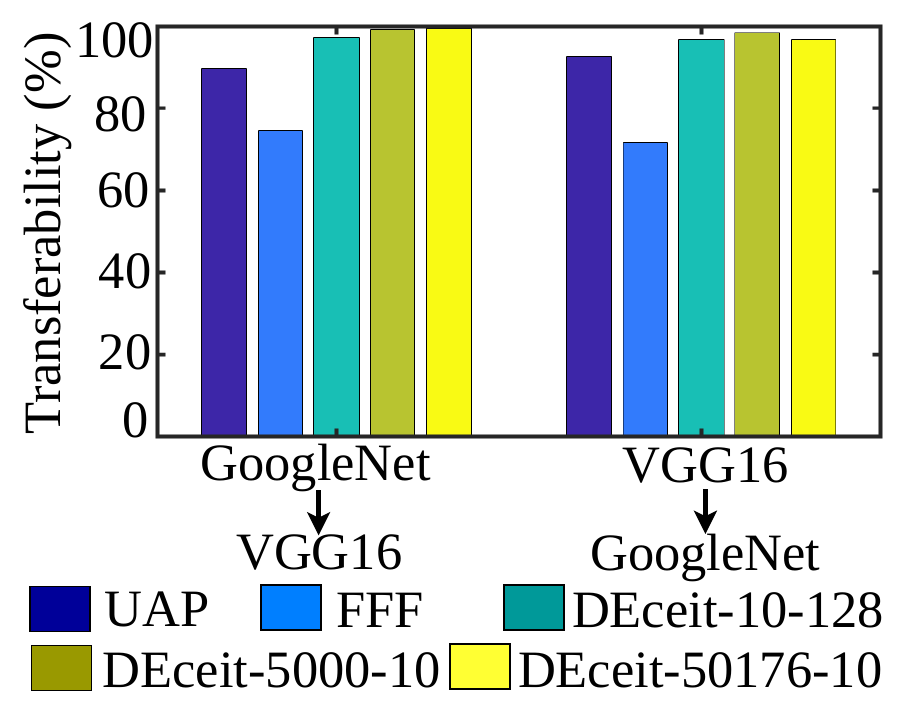}} \hspace{25pt}
    \subfigure[\label{fig:vggIv3}]{\includegraphics[width=0.35\linewidth]{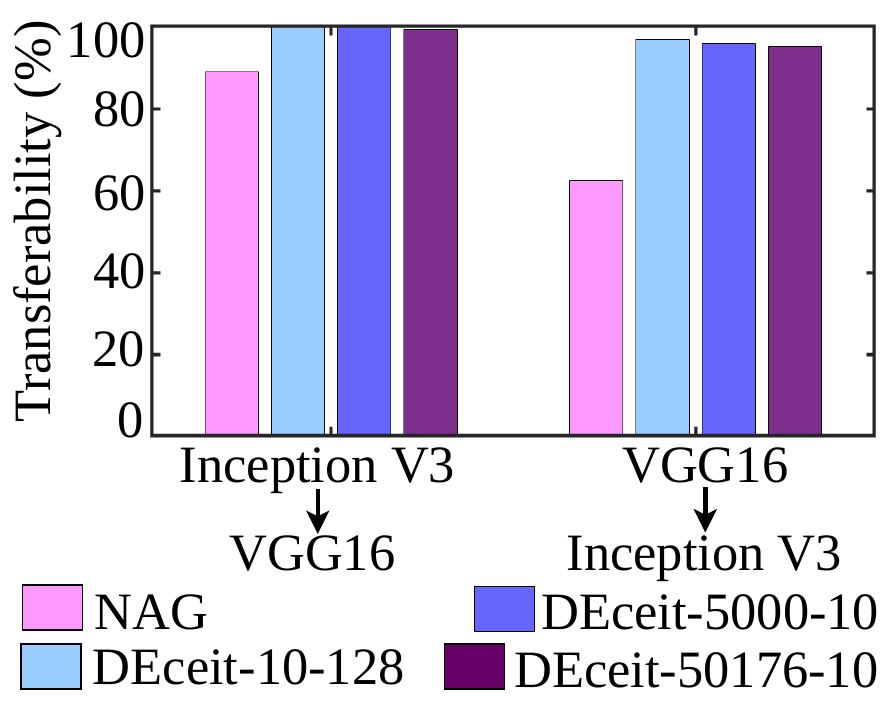}}
    \caption[Transferability of universal perturbation]{Transferability of universal perturbation obtained by DEceit on VGG16, GoogleNet and Inception V3 in comparison to the state-of-the-art. The DEceit variants are named as DEceit-$\rho$-$\beta$. (a) Better transferability is achieved by DEceit when compared with UAP and FFF on VGG16 and GoogleNet. (b) DEceit demonstrates better transferability on VGG16 and Inception V3 compared to NAG.}
    \label{fig:transferability}
\end{figure}

\subsection{Universal Adversarial Attacks}\label{sec:UAA}
In this section, we first compare the attack performance of DEceit against contending UAA algorithms. Moreover, we further discuss the transferability property of the perturbation generated by DEceit. We further investigate the effectiveness of DEceit by analyzing the effects of the generated universal perturbation on the output of the target network. Finally, we validate the efficacy of DEceit to withstand the various defense methods. 

\subsubsection{Universal Adversarial Attack Performance of DEceit}

We compare DEceit with the state-of-the-art universal non-targeted attack methods namely UAP \cite{moosavi2017universal}, FFF \cite{mopuri-bmvc-2017}, NAG \cite{Mopuri_2018_CVPR}, and GAP \cite{poursaeed2018generative} in Table \ref{tab:uapRes} (the results of the contenders are quoted from their corresponding articles). We observe from Table \ref{tab:uapRes} that the proposed method outperforms the competitors on VGG16 and GoogleNet while being the second-best on Inception V3 in terms of both Fooling Rate and PSNR when all pixels are perturbed with identical $\beta$. This indicates that DEceit is indeed capable of achieving competitive performance. Moreover, a qualitative comparison of the generated universal perturbations for UAP, FFF, and DEceit is presented in Figure \ref{fig:visual}. It is evident from Figure \ref{fig:visual} that compared to UAP and FFF the two benchmark white-box UAA techniques, the adversarial examples generated by the black-box method DEceit are minimally distorted from their corresponding original images, even when all pixels are allowed to be perturbed. From Table \ref{tab:uapRes} and Figure \ref{fig:visual}, we observe that with increasing sparsity, the Fooling Rate decreases while the PSNR improves. However, DEceit provides a perturbation of commendable quality while preserving the visual similarity by only manipulating about 10\% of the total pixels.

\subsubsection{Transferability of the universal adversarial perturbation generated by DEceit}
To further evaluate the robustness of DEceit compared to its contenders, we conduct a transferability study. Let us consider a universal adversarial perturbation $\Delta_{A}$ which is optimized for classifier $A$ and achieves a Fooling Rate of $a$ on a given set of images $\mathcal{S}$. If the same perturbation $\Delta_{A}$ is tested on another classifier $B$ on which it attains a Fooling Rate of $b$ on the same set of images $\mathcal{S}$ then the transferability from $A \rightarrow B$ is given as $\frac{b}{a} \times 100 \%$. From Figure \ref{fig:transferability} we observe that all three variants of DEceit achieve better transferability than the other methods on both classifiers, demonstrating the higher robustness of the proposed method. Interestingly, the perturbation generated by NAG which performs well on Inception V3 (as observed in Table \ref{tab:uapRes}) fails on VGG16 in Figure \ref{fig:vggIv3}, indicating its lower robustness compared to DEceit. Moreover, the transferability of DEceit only slightly changes with the choices of $\rho$, indicating resilience to the extent of sparsity. Such improved robustness of DEceit possibly arises due to its black-box setting and the efficient search capability.

\subsubsection{Analysing the Universal Adversarial Perturbations Generated by DEceit}
\begin{figure}[!ht]
    \centering
    \subfigure[\label{fig:gc1}]{\includegraphics[width=0.15\linewidth]{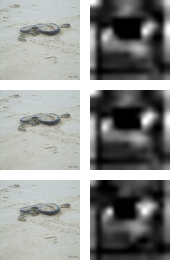}}
    \subfigure[\label{fig:gc2}]{\includegraphics[width=0.15\linewidth]{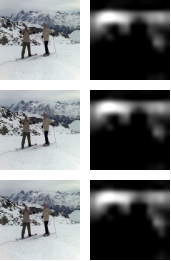}}
    \subfigure[\label{fig:gc3}]{\includegraphics[width=0.15\linewidth]{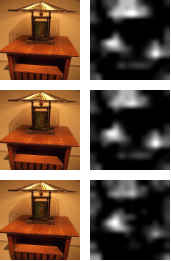}}
    \subfigure[\label{fig:gc4}]{\includegraphics[width=0.15\linewidth]{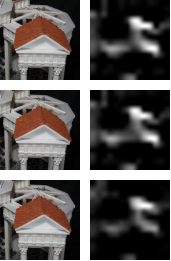}}
    \subfigure[\label{fig:gc5}]{\includegraphics[width=0.15\linewidth]{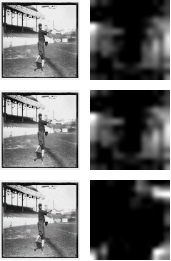}}
    \subfigure[\label{fig:gc6}]{\includegraphics[width=0.15\linewidth]{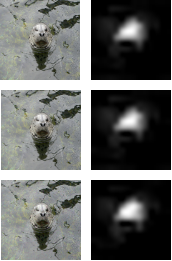}}
    \caption{Inspecting the effect of universal perturbation generated by UAA on VGG16. We consider six examples all of which successfully alter the VGG16 prediction after applying universal perturbation generated by DEceit both for $\rho=10$ and $\rho=50176$. We illustrate the six samples in sub-figures (a)-(f). In each sub-figure the left column is the input image and the right column is the corresponding GradCam {\protect\cite{selvaraju2017grad}} output. In each sub-figure respectively from top we show the original input, the input perturbed with $\rho=10$ and $\beta=128$, and the input perturbed with $\rho=50176$ and $\beta=10$.}
    \label{fig:gradcams}
\end{figure}

To better understand how DEceit manages to provide its commendable performance we employ the GradCam \cite{selvaraju2017grad} visualization tool. In Figure \ref{fig:gradcams} we consider six images that are capable to fool VGG16 network after applying the universal perturbations generated by DEceit. Specifically, in each of Figures \ref{fig:gc1}-\ref{fig:gc6} we provide a visual comparison of the GradCam outputs of the original input, input perturbed with a sparsity of $\rho=10$, and input altered by a non-sparse perturbation. A closer inspection of Figure \ref{fig:gradcams} reveals that a universal perturbation with high sparsity is still capable to alter the GradCam outputs for a diverse set of input images. In most of the cases when $\rho=10$ the visual changes in GradCam outputs are only minute indicating that the feature space representation remains almost similar after applying a highly sparse perturbation. Hence, even with small changes in the feature space representation, it is possible for the network to predict a different class for the perturbed image. This in consequence validates the significance of sparse universal adversarial attacks. Evidently, when we allow all the pixels to be perturbed the changes in GradCam outputs are more visually apparent. Further, from Figure \ref{fig:gc5}-\ref{fig:gc6} we can see that the extent of change in GradCam outputs for the same universal perturbation depends also on the input image. Thus, while designing a universal adversarial attack a perturbation may not need to drastically alter the feature representation for all the classes and can still be able to fool the classifier. This may be explained by the fact that some classes of a trained classifier may have a higher generalization error than others that in consequence make them easier to fool. This observation also validates the efficacy of the objective function used in DEceit that only maximizes the number of successful attacks.

\subsubsection{Universal Adversarial Attack Performance of DEceit Against Defense Strategies}

\begin{table}[!ht]
    \centering
    \scriptsize
    \caption{The Fooling Rate of the universal adversarial perturbation optimized by DEceit after applying image processing based defense techniques.}
    \label{tab:defenseRes}
    \vspace{5pt}
    \begin{tabular}{c|cc|cc} \toprule
        \multirow{3}{*}{defense strategies} & \multicolumn{4}{c}{Fooling Rate (\%) of UAA optimized by DEceit} \\ \cmidrule{2-5}
        & \multicolumn{2}{c|}{VGG16} & \multicolumn{2}{c}{Inception V3} \\
        & $\rho = 10$ & $\rho = 50176$ & $\rho = 10$ & $\rho = 50176$ \\ 
        & $\beta = 128$ & $\beta=10$ & $\beta = 128$ & $\beta=10$ \\ \midrule
        Without defense	&	46.79	&	83.96	&	44.23	&	83.03	\\ \midrule
        Wavelet denoising	&	41.23	&	76.88	&	39.14	&	75.21	\\
        JPEG compression 50\%	&	42.69	&	71.65	&	40.25	&	71.39	\\
        JPEG compression 75\%	&	41.12	&	70.52	&	40.64	&	70.52	\\
        Total variation denoising	&	38.55	&	72.14	&	36.92	&	70.34	\\
        Non-local mean denoising	&	39.68	&	71.91	&	38.06	&	72.16	\\
        Bilateral denoising &	41.07	&	75.23	&	40.08	&	74.57	\\
        Median filtering	&	37.16	&	77.95	&	37.55	&	77.22	\\
        Gaussian blurring &	44.35	&	76.87	&	41.69	&	74.97	\\ \bottomrule
    \end{tabular}
\end{table}

An adversarial attack is performed by perturbing the original input image by some noise. Evidently, an intuitive defense strategy can be mitigating the effect of noise by some image processing-based denoising techniques \cite{mopuri2018generalizable}. DEceit being a universal black-box attack method we do not consider the defense techniques which may include modification to the training algorithm \cite{tramer2017ensemble}, make any architectural changes \cite{metzen2017detecting}, or techniques \cite{akhtar2018defense} that are previously shown to perform poorly \cite{mopuri2018generalizable} compared to image denoising methods. Thus, in this section, we limit ourselves in validating the ability of the universal adversarial perturbation generated by DEceit to withstand the effect of image denoising-based defenses. In Table \ref{tab:defenseRes} we compare the Fooling Rate of perturbation generated by DEceit before and after applying eight denoising methods namely, wavelet \cite{chang2000adaptive}, JPEG compression (50\% and 75\%) \cite{dziugaite2016study}, total variation \cite{chambolle2004algorithm}, non-local mean \cite{buades2005non}, bilateral \cite{tomasi1998bilateral}, median filter, and Gaussian blurring. A closer inspection of Table \ref{tab:defenseRes} reveals that on both deep classifiers median filters, total-variation, and non-local mean based denoising techniques can provide effective defense when the attack is sparse with a high degree of perturbation i.e. $\rho$ and $\beta$ are respectively 10 and 128. This may be explained by the nature of such denoising techniques that attempts to remove or suppress the effect of spurious extreme values. However, when all pixels are perturbed with a lower degree of noise besides total-variation smoothing the Gaussian, bilateral, and non-local mean-based denoising methods also offer good performance on both deep networks. Further, JPEG compression provides sufficient defense against adversarial attacks as previously indicated by \cite{dziugaite2016study} even though a high level of compression may contain the risk of distorting the input image. Moreover, the efficacy of defense strategies improves on Inception V3 attesting to the better susceptibility of the classifier against adversarial attacks. Furthermore, none of the defense strategies manages to completely neutralize the effect of the universal perturbation generated by DEceit as the Fooling Rate always remains higher than the reported error rate of the classifiers under concern. Therefore, we may conclude that DEceit can generate black-box universal adversarial attacks that are mostly robust against the defense methods.

\begin{table}[!ht]
    \centering
    \scriptsize
    \caption{Experimental Protocol for IDA}
    \label{IDAprotocol}
    \vspace{5pt}
    \begin{tabular}{p{35mm}p{110mm}}
    \toprule
    Algorithm & Experimental Protocol \\ \midrule
       PBBA and GAP  & DEceit are run on the entire ImageNet 2012 validation set containing 50000 images. \\ \midrule
       LogBarrier & We compare the performance of DEceit on 500 randomly selected images following the original article. The performance of DEceit is averaged over 5 such subsets of images for the sake of fairness. \\ \midrule
       SIMBA & We compare the performance of DEceit on 1000 randomly selected images following the original article. The performance of DEceit is averaged over 5 such subsets of images for the sake of fairness.\\ \midrule
       GenAttack &  We compare the performance of DEceit on 100 correctly classified images following the original article. The performance of DEceit is averaged over 5 such subsets of images for the sake of fairness.\\ \midrule
       SparseFool and OPA & Performances are compared in terms of average performance on 5 subsets each containing 2000 correctly classified images. The same 5 subsets of 2000 images are used for all three methods including DEceit. \\ \bottomrule
    \end{tabular}
\end{table}

\begin{table*}[!ht]
    \centering
    \caption{Comparison of DEceit with non-targeted IDA methods on ImageNet 2012 validation set.}
    \label{tab:idaRes}
    \vspace{5pt}
    \begin{threeparttable}
    \scriptsize
    \begin{tabular}{ccccc|cc|cc} \toprule
    & & & & & \multicolumn{2}{c|}{Contender} & \multicolumn{2}{c}{DEceit (Ours)} \\ \cmidrule{6-9}
    Method & $\rho$ & $\beta$ & Attacked & Number of & Average & Fooling & Average & Fooling \\
    & & & Network & Images & Query & Rate (\%) & Query & Rate (\%) \\ \midrule
    PBBA & 50176 & 0.05 & Inception V3 & 50000 & 722 & 98.5 & \textbf{620} & \textbf{98.79} \\
    GAP & 50176 & 10 & Inception V3 & 50000 & - & 98.3 & - & \textbf{100} \\
    LogBarrier & 50176 & 8 & ResNet 50 & 500 & - & 95.2 & - & \textbf{99.1} \\
    SIMBA & 50176 & 3 & Inception V3 & 1000 & 1284 & 97.8 & \textbf{650} & \textbf{98.1}\\ 
    GenAttack & 50176 & 0.05 & Inception V3 & 100 & 11081 & \textbf{100} & \textbf{535} & \textbf{100}\\
    SparseFool & 7024$^{*}$ & 255 & Inception V3 & 2000 & - & \textbf{100} & - & \textbf{100} \\
    OPA & 1 & 128 & Inception V3 & 2000 & - & 37.25 & - & \textbf{46.12} \\
    \bottomrule
    \end{tabular}
    \begin{tablenotes}
    \item ''-" indicates that the Average Query has not been reported by the contender for comparison. 
    \item $*$: During the comparison with SparseFool 7024 is the average number of perturbed pixels.
    \item The best results are boldfaced.
    \end{tablenotes}
    \end{threeparttable}
\end{table*}

\subsection{Image-Dependent Attacks}\label{sec:IDA}
DEceit can directly be applied for IDA by feeding $\mathcal{S}$ as a singleton set containing only the image to be perturbed. We evaluate the efficacy of DEceit for IDA by comparing its performance with the state-of-the-art methods namely PBBA \cite{pmlr-v97-moon19a}, GAP \cite{poursaeed2018generative}, LogBarrier \cite{Finlay_2019_ICCV}, SIMBA \cite{shiva2017simple}, GenAttack \cite{alzanot2019genattack}, SparseFool \cite{Modas_2019_CVPR}, and OPA \cite{su2019one}. During the comparison of DEceit with IDA techniques on the ImageNet 2012 validation set the experimental protocol followed by us is detailed in Table \ref{IDAprotocol}. In all cases, the number of perturbed pixels $\rho$ and the range of allowable distortion $\beta$ for DEceit are kept identical to the corresponding contender.

We can observe from Table \ref{tab:idaRes} that DEceit outperforms PBBA, GAP, LogBarrier, SIMBA, and OPA in terms of Fooling Rate. Furthermore, DEceit requires a lesser number of average queries compared to PBBA and SIMBA supporting its efficiency in performing an economic IDA. Interestingly, DEceit performs equivalently with GenAttack and SparseFool. However, GenAttack requires a higher number of average queries which may be due to its inability to handle high-dimensional optimization problems. Whereas, SparseFool though capable of performing while incurring a slightly lower computational cost is neither a black-box technique nor can be directly extended to UAA, making DEceit an attractive alternative. 

\begin{figure}[!ht]
    \centering
    \subfigure[\label{fig:timings}]{\includegraphics[width=0.4\linewidth]{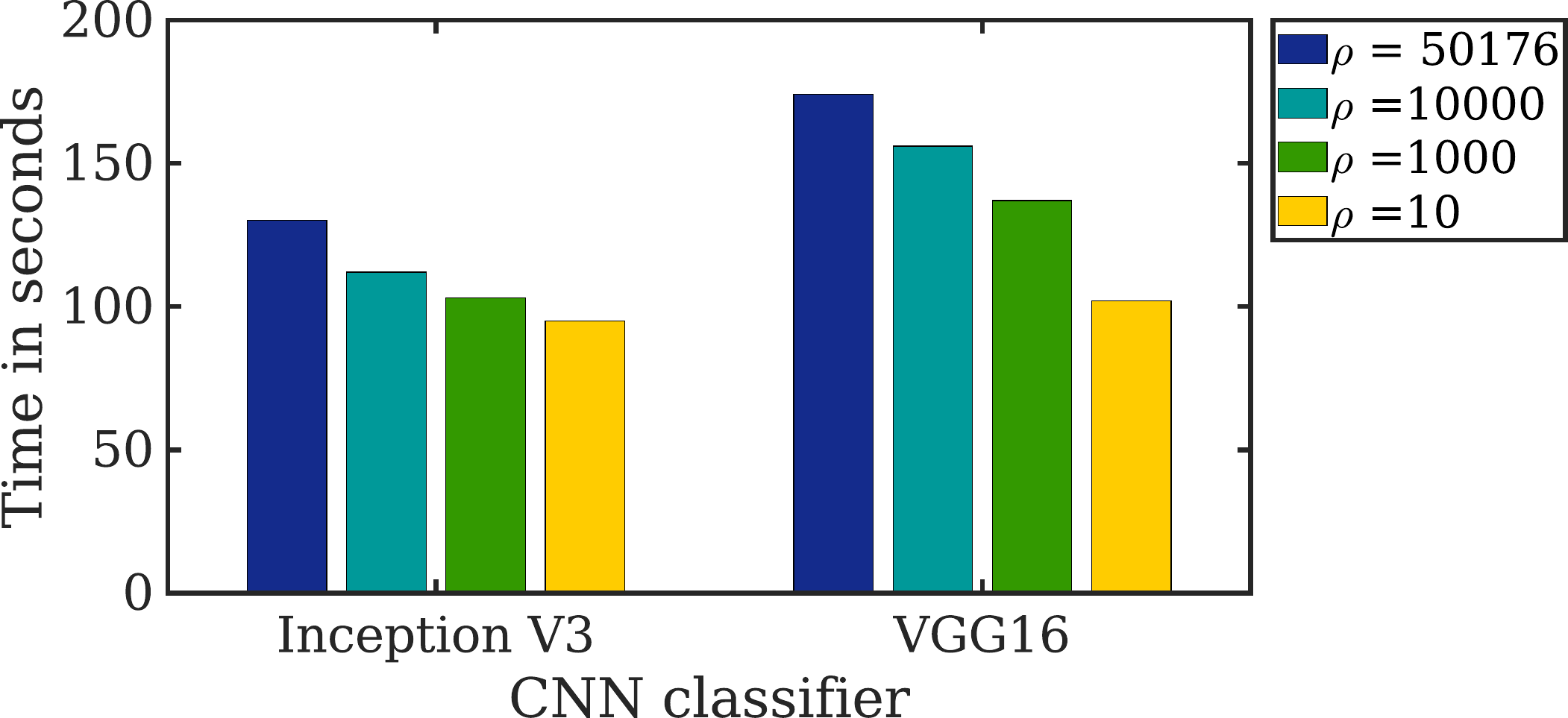}}
    \subfigure[\label{fig:fes}]{\includegraphics[width=0.4\linewidth]{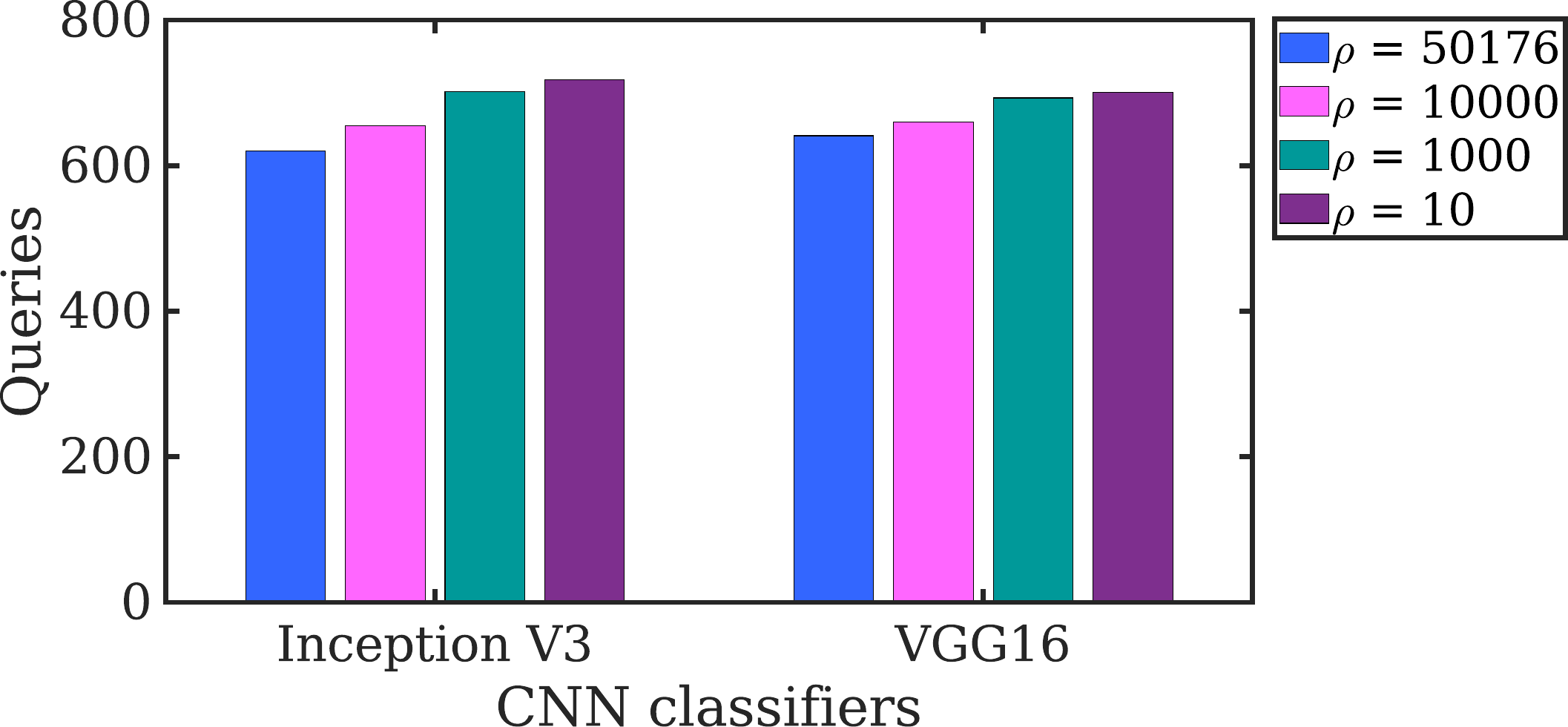}}
    \subfigure[\label{fig:normalized_timings}]{\includegraphics[width=0.4\linewidth]{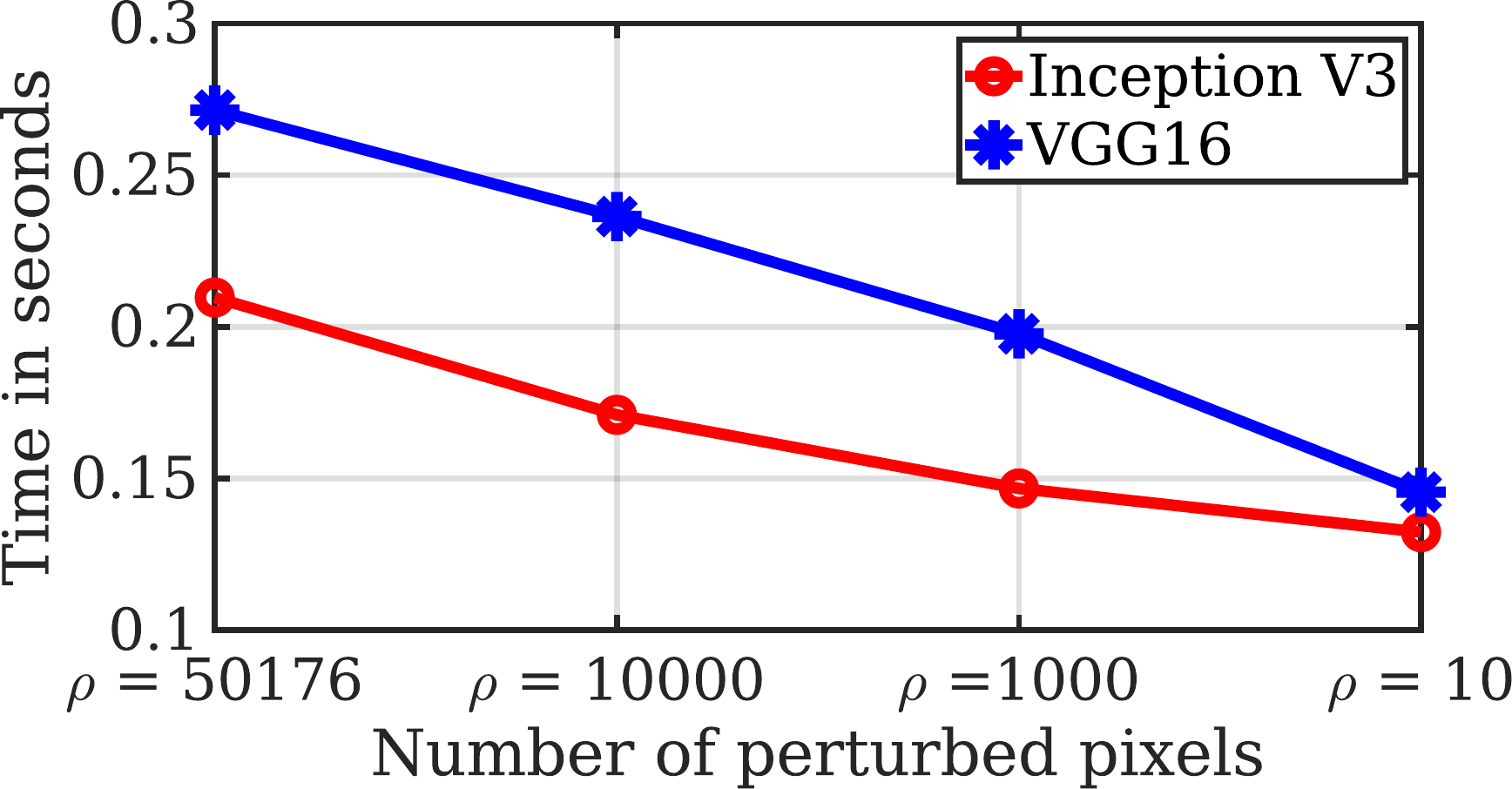}}
    \caption{Computational cost of DEceit in practice. (a) The average time in seconds taken by DEceit to perform a successful IDA for Inception V3 and VGG16. (b) The average number of queries required by DEceit to perform a successful IDA for Inception V3 and VGG16. (b) The average time in seconds taken by DEceit to run a single optimization step for different choices of $\rho$ for Inception V3 and VGG!6.}
    \label{fig:timing_record}
\end{figure}

The lower computational cost incurred by DEceit in practice can be further highlighted by Figure \ref{fig:timing_record}. From Figure \ref{fig:timings} we can observe that irrespective of the choice of CNN classifier with the increasing number of pixels to be perturbed the time taken by DEceit also increases. Even when we allow all pixels to be perturbed DEceit takes at most 174 seconds to perform a successful IDA on VGG16. If $\rho$ is set to 10, then DEceit takes the lowest average time of 95 seconds to generate an effective image-dependent perturbation. Thus, DEceit can provide an economical solution to the IDA problem in practice. However, decreasing $\rho$ consequently increases the difficulty of the optimization problem, as we have to find an effective perturbation by modifying only a fewer number of pixels. Even though, we attempt to compensate for the sparsity with a higher degree of allowable perturbation we may still require a larger number of queries in practice. This becomes apparent from Figure \ref{fig:fes} where with increasing sparsity the average number of queries required to perform a successful attack also increases. Therefore, to know how the sparsity actually impacts the computational cost of DEceit we need to normalize the average time by the average number of queries as in Figure \ref{fig:normalized_timings}. From Figure \ref{fig:normalized_timings} we can validate that the time taken by DEceit to complete a single optimization step is indeed dependent on the choice of $\rho$. This observation is in line with the theoretical analysis of DEceit described in Theorem \ref{timeCompTheo} which shows that the computational complexity of the proposed method is linearly proportional to $\rho$. In this experiment, all the timings are averaged over 1000 successful IDA performed by DEceit where the algorithm is run on a Titan XP GPU. 

\section{Conclusion and future works}\label{sec:conclu}
In this article, we introduced DEceit, a black-box pixel-restricted, non-targeted, universal/image-dependent adversarial perturbation technique. We expressed the task of finding an adversarial perturbation as a constrained optimization problem. To effectively solve this real-valued high-dimensional optimization problem, we devised DEceit by modifying DE by incorporating a novel mix of scale-factor and mutation switching strategies. Finally, we empirically validated the capability of DEceit to perform effective, imperceptible, and robust adversarial attacks.

DEceit may be extended by focusing on the contradictory objectives of maximizing the effectiveness of a perturbation while minimizing the visual distortions, which can be expressed as a multi-objective optimization problem, solvable by a tailored DE variant. DEceit may also benefit from searching a perturbation in a transformed space instead of the native image space \cite{guo19simba}. Another interesting direction of research for DEceit type adversarial attack algorithms can be investigating how such techniques actually work in practice. Such analysis may be proven extremely useful for designing tailored defense techniques as well as improve the robustness of CNN training algorithms. Further, the mutation scale parameter and the crossover probability in DE type optimizers are not easy to properly tune. In DEceit we solve this problem by an efficient and intuitive heuristic. However, as both the operations are linear in nature they can be integrated (after some minimal alterations) in a network structure where the associated parameters can be learned through back-propagation of error. The primary concern with this solution is to appropriately modify the selection stage to make it compatible with the gradient propagation in the neural network training procedure. Further, the objective function needs to be expressed in a differentiable form. Thus, an interesting scope of future extension may also include designing a DE variant that can incorporate the benefits of learning with its commendable searching capability.

\section*{Acknowledgement}
We are grateful to the NVIDIA Corporation for donating us the Titan XP GPU used for this research.

\bibliography{main}

\end{document}